\documentclass[10pt,twocolumn]{IEEEtran}

\usepackage{amsfonts}
\usepackage{subfigure}
\usepackage{amsmath}
\usepackage{amssymb}
\usepackage{epsfig}
\usepackage{mathrsfs}
\usepackage{graphicx}
\usepackage{mathrsfs}
\usepackage{dsfont}
\usepackage{algorithm}
\usepackage{caption}
\usepackage{algpseudocode}
\usepackage{color}

\DeclareMathAlphabet{\baz}{OML}{cmm}{b}{i}

\def\bA{\mbox{\boldmath $A$}}
\def\bB{\mbox{\boldmath $B$}}
\def\bC{\mbox{\boldmath $C$}}
\def\bD{\mbox{\boldmath $D$}}

\def\bF{\mbox{\boldmath $F$}}
\def\bG{\mbox{\boldmath $G$}}
\def\bH{\mbox{\boldmath $H$}}
\def\bI{\mbox{\boldmath $I$}}

\def\bM{\mbox{\boldmath $M$}}
\def\bP{\mbox{\boldmath $P$}}
\def\bQ{\mbox{\boldmath $Q$}}
\def\bR{\mbox{\boldmath $R$}}
\def\bS{\mbox{\boldmath $S$}}
\def\bT{\mbox{\boldmath $T$}}
\def\bU{\mbox{\boldmath $U$}}
\def\bV{\mbox{\boldmath $V$}}
\def\bW{\mbox{\boldmath $W$}}
\def\bX{\mbox{\boldmath $X$}}

\def\b0{\mbox{\boldmath $0$}}

\def\bee{\mbox{\boldmath $e$}}
\def\bff{\mbox{\boldmath $f$}}
\def\bg{\mbox{\boldmath $g$}}

\def\br{\mbox{\boldmath $r$}}
\def\bs{\mbox{\boldmath $s$}}
\def\bt{\mbox{\boldmath $t$}}
\def\bu{\mbox{\boldmath $u$}}

\def\bx{\baz{x}}

\def\by{\mbox{\boldmath $y$}}
\def\bz{\mbox{\boldmath $z$}}

\def\bmu{\mbox{\boldmath $\mu$}}

\def\bv{\mbox{\boldmath $v$}}

\newenvironment{proof}[1][Proof]{\noindent \textbf{#1.} }{\qedsymbol}
\newcommand{\qedsymbol}{\hspace{\fill}\rule{1.5ex}{1.5ex}}

\title{Diffusion Adaptation Strategies for Distributed Estimation over Gaussian Markov Random Fields}

\author{Paolo~Di Lorenzo,~\IEEEmembership{Member,~IEEE}\\
Department of Information, Electronics, and Telecommunications \\ ``Sapienza'' University of Rome, Via Eudossiana 18, 00184 Rome, Italy.\\ e-mail: {\tt \{dilorenzo\}@infocom.uniroma1.it}
\thanks{This work has been supported by TROPIC Project, Nr. 318784.}}

\begin{document}

\maketitle

\begin{abstract}
The aim of this paper is to propose diffusion strategies for distributed estimation over adaptive networks,
assuming the presence of spatially correlated measurements distributed according to a Gaussian Markov random field (GMRF) model.
The proposed methods incorporate prior information about the statistical dependency among observations, while at the same time processing
data in real-time and in a fully decentralized manner. A detailed mean-square analysis is carried out in order to prove stability and evaluate
the steady-state performance of the proposed strategies. Finally, we also illustrate how the proposed techniques can be easily extended in
order to incorporate thresholding operators for sparsity recovery applications. Numerical results show the potential advantages of using such
techniques for distributed learning in adaptive networks deployed over GMRF.
\end{abstract}

\begin{keywords}
Distributed LMS estimation, adaptive networks, correlated noise, Gaussian Markov random fields, sparse adaptive estimation, sparse vector.
\end{keywords}

\section{Introduction}


We consider the problem of distributed estimation \cite{Barb-Sard-Dilo}, where a set of nodes is required to collectively estimate some vector parameter of interest from noisy measurements by relying solely on in-network processing. We consider an ad-hoc network consisting of $N$ nodes that are distributed over some geographic region. At every time instant $k$, every node $i$ collects a scalar measurement $x_i[k]$ and a $1\times M$ regression vector $\bu_{i}[k]$. The objective is for the nodes in the network to use the collected data $\{x_i[k],\bu_i[k]\}$ to estimate some $M\times 1$ parameter vector $\boldsymbol{\theta}_0$ in a distributed manner. There are a couple of distributed strategies that have been developed in the literature for such purposes. One typical strategy is the incremental approach \cite{Bertsekas2}-\hspace{-.1mm}\cite{Li-Chambers-Lopes-Sayed}, where each node communicates only with one neighbor at a time over a cyclic path.
However, determining a cyclic path that covers all nodes is an NP-hard problem \cite{Karp} and, in addition, cyclic trajectories are prone
to link and node failures.
To address these difficulties, consensus based \cite{Schizas-Mateos-Giannakis} and diffusion-based \cite{Lopes_Sayed,Cattivelli_Sayed}
techniques were proposed and studied in literature. In these implementations, the nodes exchange information locally and cooperate with
each other without the need for a central processor. In this way, information is processed on the fly and the data diffuse across the
network by means of a real-time sharing mechanism. Since diffusion strategies have shown to be more stable and performing than consensus
networks \cite{Tu-Sayed2}, we will focus our attention on diffusion type of networks. In view of their robustness and adaptation properties,
diffusion networks have been applied to model a variety of self-organized behavior encountered in nature, such as birds flying in formation
\cite{Cattivelli-Sayed3}, fish foraging for food \cite{Tu-Sayed} or bacteria motility \cite{Chen-Zhao-Sayed}. Diffusion adaptation has
also been used for distributed optimization and learning \cite{Chen-Sayed}, to solve dynamic resource allocation problems in
cognitive radios \cite{Dilo-Barb-Sayed} and distributed spectrum estimation in small cell networks \cite{Dilorenzo-Barbarossa-Sayed2},
to perform robust system modeling  \cite{Chouvardas-Slavakis-Theodoridis}, and to implement distributed learning over mixture models in
pattern recognition applications \cite{Towfic-Chen-Sayed}.

A characteristic of the observed signal that can be advantageously exploited to improve the estimation accuracy is the sparsity of the
parameter to be estimated, i.e., the vector $\boldsymbol{\theta}_0$ contains only a few relatively large coefficients among many negligible
ones. Any prior information about the sparsity of $\boldsymbol{\theta}_0$ can be exploited to help improve the estimation performance, as
demonstrated in many recent efforts in the area of compressive sensing (CS) \cite{Donoho}-\cite{Baraniuk}. Up to now, most CS efforts have
concentrated on batch recovery methods, where the estimation of the desired vector is achieved from a collection of a fixed number of
measurements. In this paper, we are instead interested in adaptive techniques that allow the recovery of sparse vectors to be pursued both
recursively and distributively as new data arrive at the nodes.
Such schemes are useful in several contexts such as in the analysis of prostate cancer data \cite{Tibshirani},
\cite{Mateos-Bazerque-Giannakis}, spectrum sensing in cognitive radio \cite{Bazerque-Giannakis},\cite{Dilorenzo-Barbarossa-Sayed2},
and spectrum estimation in wireless sensor networks \cite{Schizas-Mateos-Giannakis}.
Motivated by the LASSO technique \cite{Tibshirani} and by connections with compressive sensing  \cite{Donoho}-\cite{Baraniuk}, several algorithms for sparse adaptive filtering have been proposed based on Least Mean Squares (LMS) \cite{Chen-Gu-Hero}-\cite{Gu-Jin-Mei}, Recursive Least Squares (RLS) \cite{Angelosante-Bazerque-Giannakis,Babadi-Kalouptisidis-Tarokh}, projection-based methods \cite{Kopsinis-Slavakis-Theodoridis}, and thresholding operators \cite{Slavakis-Kopsinis-Theodoridis-McLaughlin}.
A couple of distributed algorithms implementing LASSO over ad-hoc networks have also been considered before, although their main purpose has
been to use the network to solve a {\em batch} processing problem \cite{Mateos-Bazerque-Giannakis,Mota-Xavier-Aguiar-Puschel}.
One basic idea in all these previously developed sparsity-aware techniques is to introduce a convex penalty term into the cost function to
favor sparsity. Our purpose in this work is to use both {\em adaptive} and {\em distributed} solutions that are able to exploit and track sparsity
while at the same time processing data in real-time and in a fully decentralized manner. Doing so would endow networks with learning
abilities and would allow them to learn the sparse structure from the incoming data recursively and, therefore, to {\em track} variations
in the sparsity pattern of the underlying vector as well. Investigations on sparsity-aware, adaptive, and distributed solutions appear in \cite{Chouvardas-Slavakis-Kopsinis-Theodoridis}-\cite{Liu-Liu-Li}.
In \cite{Liu-Li-Zhang}-\cite{DiLorenzo-Sayed}, the authors employed diffusion techniques that are able to identify and track sparsity over
networks in a distributed manner thanks to the use of convex regularization terms. In the related work
\cite{Chouvardas-Slavakis-Kopsinis-Theodoridis}, the authors employ projection techniques onto hyperslabs and weighted $\ell_1$
balls to develop a useful sparsity-aware algorithm for distributed learning over diffusion networks. Sparse distributed recursive least squares
solutions were also proposed in \cite{Sardellitti-Barbarossa}-\cite{Liu-Liu-Li}.

All the previous methods considered the simple situation where the observations are statistically independent.
In some circumstances, however, this assumption is unjustified. This is the case, for example, when the sensors monitor a
field of spatially correlated values, like a temperature or atmospheric pressure field. In such cases, nearby nodes sense
correlated values and then the statistical independence assumption is no longer valid. It is then of interest, in such cases,
to check whether the statistical properties of the observations can still induce a structure on the joint probability density function
(pdf) that can be exploited to improve network efficiency. There is indeed a broad class of observation models where the joint pdf cannot
be factorized into the product of the individual pdf's pertaining to each node, but it can still be factorized into functions of subsets
of variables. For instance, this is the case of Markov random fields and Bayes networks \cite{Lauritzen-book}.  A natural
approach in these settings is to incorporate additional prior knowledge in the form of structure and/or sparsity in the relationships
among observations. In particular, graphical models provide a very useful method of representing the structure of conditional dependence
among random variables through the use of graphs \cite{Lauritzen-book}. In the Gaussian case, this structure leads to sparsity in
the inverse covariance matrix and allows for efficient implementation of statistical inference algorithms, e.g., belief propagation.
Several techniques have been proposed in the literature for covariance estimation, where the structure of the dependency graph is assumed
to be known, and covariance selection, where also the structure of the graph is unknown and must be inferred from measurements
(see, e.g., \cite{Wiesel-Hero}, \cite{Meinhausen-Buhlmann} and references therein). Recent works on distributed estimation over GMRF appear in
\cite{Barb-Sard-Dilo}, \cite{Dog-Liu}-\cite{Fang-Li}.

The contribution of this paper is threefold: (a) The development of novel distributed LMS strategies for adaptive estimation over networks,
which are able to exploit prior knowledge regarding the spatial correlation among nodes observations distributed according to a GMRF
(To the best of our knowledge this is the first strategy proposed in the literature that exploits the spatial correlation among data
in an adaptive and distributed fashion);  (b) The derivation of a detailed mean-square analysis that provides closed form
expressions for the mean-square deviation (MSD) achieved at convergence by the proposed strategies; (c) The extension of the proposed strategies to include thresholding
operators, which endow the algorithms of powerful sparsity recovery capabilities.

The paper is organized as follows. In section II we recall some basic notions from GMRF that will be largely used throughout the paper.
In section III we develop diffusion LMS strategies for distributed estimation over adaptive networks, considering spatially correlated
observations among nodes. Section IV provides a detailed performance analysis, which includes mean stability and mean-square performance.
In Section V, we extend the previous strategies in order to improve performance under sparsity of the vector to be estimated. Section VI
provides simulation results in support of the theoretical analysis. Finally, section VII draws some conclusions and possible future
developments.

\section{Gaussian Markov Random Fields}

In this section, we briefly recall basic notions from the theory of Gaussian Markov random fields, as this will form the basis of the distributed estimation algorithms developed in the ensuing sections.

A Markov random field is represented through an undirected graph. More specifically, a Markov network consists of \cite{Lauritzen-book}:
\begin{enumerate}
\item An undirected graph $G_{sd} = (V_{sd},E_{sd})$, where each vertex $v \in V_{sd}$ represents a
random variable and each edge $\{u,v\} \in E_{sd}$ represents conditional statistical dependency
between the random variables $u$ and $v$;
\item
A set of potential (or compatibility) functions $\psi_c(\bx_c)$ (also called clique potentials),
that associate a non-negative number to the cliques \footnote{A clique is a subset
of nodes which are fully connected and maximal, i.e. no additional node can be added
to the subset so that the subset remains fully connected.} of $G_{sd}$.
\end{enumerate}
Let us denote by ${\cal C}$ the set of all cliques present in the graph. The random vector $\bx$ is
Markovian if its joint pdf admits the following factorization
\begin{equation}
\label{Markov_factor}
p(\bx)=\frac{1}{Z}\prod_{c\in {\cal C}}\psi_c(\bx_c),
\end{equation}
where $\bx_c$ denotes the vector of variables belonging
to the clique $c$. The functions $\psi_c(\bx_c)$ are called {\it compatibility functions}.
The term $Z$ is simply a normalization factor necessary to guarantee that
$p(\bx)$ is a valid pdf.
A node $p$ is conditionally independent of another node $q$ in the Markov network,
given some set $S$ of nodes, if every path from $p$ to $q$ passes through a
node in $S$. Hence, representing a set of random variables
by drawing the correspondent Markov graph is a meaningful pictorial
way to identify the conditional dependencies occurring across the random variables.
If the product in (\ref{Markov_factor}) is strictly positive for any $\bx$, we can introduce the functions
\begin{equation}
V_c(\bx_c)=-\log \psi_c(\bx_c)
\end{equation}
so that (\ref{Markov_factor}) can be rewritten in exponential form as
\begin{equation}
\label{Markov_factor_exp}
p(\bx)=\frac{1}{Z}\exp\left(-\sum_{c\in {\cal C}}V_c(\bx_c)\right).
\end{equation}
This distribution is known, in physics, as the Gibbs (or Boltzman) distribution with interaction
{\it potentials} $V_c(\bx_c)$ and {\it energy} $\sum_{c\in {\cal C}}V_c(\bx_c)$.\\
The dependency graph $G_{sd}$ conveys the key probabilistic information through absent
edges: If nodes $i$ and $j$ are not neighbors, the random variables $x_i$ and $x_j$ are statistically independent, conditioned to the other variables. This is the so called {\it pairwise Markov property}.
Given a subset $a \subset V_{sd}$ of vertices, $p({\bx})$ factorizes as
\begin{equation}
\label{Markov_factor_2}
p(\bx)=\frac{1}{Z}\prod_{c:c\cap a\neq\emptyset}\psi_c(\bx_c)\,\prod_{c:c\cap a = \emptyset}\psi_c(\bx_c)
\end{equation}
where the second factor does not depend on $a$. As a consequence, denoting by $S-a$ the set of all nodes except the nodes in $a$ and by
${\cal N}_a$ the set of neighbors of the nodes in $a$, $p(\bx_a/\bx_{S-a})$ reduces to $p(\bx_a/{\cal N}_a)$. Furthermore,
\begin{align}
\label{Markov_factor_3}
p(\bx_a/{\cal N}_a)\;=\;
&\frac{1}{Z_a}\prod_{c:c\cap a\neq\emptyset}\psi_c(\bx_c) \nonumber\\
=&\frac{1}{Z_a}\,\exp\left(-\sum_{c:c\cap a\neq\emptyset}\,V_c(\bx_c)\right).
\end{align}
This property states that the joint pdf factorizes in terms that contain only variables
whose vertices are neighbors. An important example of jointly Markov random variables is the Gaussian Markov
Random Field (GMRF), characterized by having a pdf expressed as in (\ref{Markov_factor_exp}), with the additional property
that the energy function is a quadratic function of the variables. In particular, a
vector $\bx$ of random variables is a GMRF if its joint pdf can be written as
\begin{align}
\label{gmrf}
p(\bx)\;=\;\sqrt{\frac{|\bB|}{(2\pi)^N }}\,e^{\displaystyle  -\frac{1}{2}(\bx-\bmu)^T\bB(\bx-\bmu)},
\end{align}
where $\bmu=\mathbb{E}\{\bx\}$ is the expected value of $\bx$, $\bB=\bC^{-1}$ is the so called {\it precision} matrix, with
$\bC=\mathbb{E}\{(\bx-\bmu)(\bx-\bmu)^T\}$ denoting the covariance matrix of $\bx$. In this case, the {\it Markovianity} of $\bx$
manifests itself through the {\it sparsity} of the precision matrix. Indeed, as a particular case of (\ref{Markov_factor_3}),
the coefficient $b_{i,j}$ of $\bB$ is different from zero if and only if nodes $i$ and $j$ are neighbors in the dependency graph, i.e.,
the corresponding random variables $x_i$ and $x_j$ are statistically dependent, conditioned to the other variables.
The following result from \cite{Anandkumar-Tong-Swami} provides an explicit expression between the coefficients of the covariance and the
precision matrices for acyclic graphs. The entries of the precision matrix $\bB=\{b_{i,j}\}$, for a positive definite covariance
matrix $\bC=\{c_{i,j}\}$ and an acyclic dependency graph, are:
\begin{align}\label{prec_cov_matrices}
b_{i,i}&=\frac{1}{c_{i,i}}+\frac{c^2_{i,j}/c_{i,i}}{c_{i,i}c_{j,j}-c^2_{i,j}}; \\
b_{i,j}&=\left\{
         \begin{array}{ll}
           \displaystyle \frac{-c_{i,j}}{c_{i,i}c_{j,j}-c^2_{i,j}},  \quad \hbox{$j\in {\cal M}_i$;}\\
           0,  \hspace{2.3cm}\hbox{o.w.}
         \end{array}
       \right.
\end{align}
Let us assume that $c_{i,i}=\sigma^2$, for all $i$, and that the amount of correlation between the neighbors $(i,j)$ of the dependency graph
is specified by an arbitrary function $g(\cdot)$, which has the Euclidean distance $d_{ij}$ as its argument, i.e. $c_{i,j}=\sigma^2 g(d_{ij})$.
Furthermore, if we assume that the function $g(\cdot)$ is a monotonically non-increasing function of the distance (since amount of correlation usually decays as nodes become farther apart) and $g(0) = \nu < 1$, exploiting a result from \cite{Anandkumar-Tong-Swami}, it holds that matrix $\bC$ is positive definite.

\section{Diffusion Adaptation over Gaussian Markov Random Fields}

Let us consider a network composed of $N$ nodes, where the observation $x_i[k]$ collected by node $i$, at time $k$, follows the linear model
\begin{equation}\label{linear_model}
x_i[k]=\bu_i^T[k]\;\boldsymbol{\theta}_0+v_i[k], \quad i=1,\ldots,N
\end{equation}
where $\boldsymbol{\theta}_0$ is the $M$-size column vector to be estimated, and $\bu_i[k]$ is a known time-varying regression vector of size $M$. The observation noise vector $\bv[k]=[v_1[k],\ldots,v_N[k]]^T$ is distributed according to a Gaussian Markov random field with zero-mean and precision matrix $\bB$. Since the vector $\bv$ is a Gaussian Markov random field, the precision matrix $\bB$ is typically a sparse matrix that reflects the structure of the dependency graph among the observed random variables.

Following a maximum likelihhod (ML) approach, the optimal estimate for $\boldsymbol{\theta}_0$ can be found as the vector that maximizes the log-likelihood function \cite{Kay}, i.e.,
as the solution of the following optimization problem:
\begin{equation}\label{MAP_est}
\max_{\boldsymbol{\theta}} \;\; \mathbb{E} \left\{ \log\left(p(\bx[k],\boldsymbol{\theta})\right)\right\}
\end{equation}
where $p(\bx[k],\boldsymbol{\theta})$ is the pdf of the observation vector $\bx[k]=[x_1[k],\ldots,x_N[k]]^T$ collected by all the nodes at time
$k$, which depends on the unknown parameter $\boldsymbol{\theta}$. Since the observation noise is distributed according to a GMRF, exploiting
the joint pdf expression (\ref{gmrf}) and the linear observation model (\ref{linear_model}), the cooperative estimation
problem in (\ref{MAP_est}) is equivalent to the minimization of the following mean-square error cost function:
\begin{equation}\label{Cent_MMSE_GMRF}
\boldsymbol{\hat{\theta}}[k]= \displaystyle {\rm arg} \min_{\boldsymbol{\theta}} \;\; \frac{1}{2}\;\mathbb{E}\left\{\|\bx[k]-\bU[k]\boldsymbol{\theta}\|^2_{\boldsymbol{B}}\right\}
\end{equation}
where $\bU[k]=\left[\bu^T_1[k]|\ldots|\bu^T_N[k]\right]$, and we have introduced the weighted norm $\|\by\|^2_{\boldsymbol{X}}=\by^T\bX\by$,
with $\bX$ denoting a generic positive definite matrix. From (\ref{Cent_MMSE_GMRF}), we have that the Gaussianity of the observations leads to the
mean-square error cost function, while the Markovianity manifests itself through the presence of the sparse precision matrix $\bB$. In the case of statistically
independent observations, the precision matrix $\bB$ in (\ref{Cent_MMSE_GMRF}) becomes a positive diagonal matrix, as already considered
in many previous works, e.g., \cite{Schizas-Mateos-Giannakis}-\cite{Cattivelli_Sayed}.

For jointly stationary $\bx[k]$ and $\bU[k]$, if the moments $\bR_{UB}=\mathbb{E}\{\bU[k]^T\bB\bU[k]\}$ and
$\bR_{UBx}=\mathbb{E}\{\bU^T[k]\bB\bx[k]\}$ were known, the optimal solution of (\ref{Cent_MMSE_GMRF}) is given by:
\begin{equation}\label{opt_sol}
\boldsymbol{\hat{\theta}}=\bR_{UB}^{-1}\bR_{UBx}.
\end{equation}
Nevertheless, in many linear regression applications involving online processing of data, the moments $\bR_{UB}$ and $\bR_{UBx}$ may be either unavailable or time varying, and thus impossible to update continuously. For this reason, adaptive solutions relying on instantaneous information are usually adopted in order to avoid the need to know the signal statistics beforehand. In general, the minimization of (\ref{Cent_MMSE_GMRF}) can be computed using a centralized algorithm, which can be run by a fusion center once all nodes transmit their data $\{x_i[k],\bu_i[k]\}$, for all $i$, to it. A centralized LMS algorithm that attempts to find the solution of problem (\ref{Cent_MMSE_GMRF}) is given by the recursion
\begin{equation}\label{cent_LMS}
\boldsymbol{\theta}[k]=\boldsymbol{\theta}[k-1]+\mu\;\bU[k]^T\bB\left(\bx[k]-\bU[k]\boldsymbol{\theta}[k-1]\right)
\end{equation}
$k\geq0$, with $\mu$ denoting a sufficiently small step-size. Such an approach calls for sufficient communications resources
to transmit the data back and forth between the nodes and the central processor, which limits the autonomy of the network,
besides adding a critical point of failure in the network due to the presence of a central node. In addition, a centralized solution may limit the ability of the nodes to adapt in real-time to time-varying statistical profiles.

In this paper our emphasis is on a distributed solution, where the nodes estimate the common parameter $\boldsymbol{\theta}_0$ by
relying solely on in-network processing through the exchange of data only between neighbors. The interaction among the nodes
is modeled as an undirected graph $G=(V_c,E_c)$ and is described by a symmetric $N\times N$ adjacency matrix $\bA:=\{a_{i,j}\}$, whose entries $a_{i,j}$
are either positive or zero, depending on wether there is a link between nodes $i$ and $j$ or not.
To ensure that the data from an arbitrary node can eventually percolate through the entire network, the following is assumed:

\noindent {\textbf{Assumption 1 (Connectivity)}}  \textit{The network graph is connected; i.e., there exists a (possibly) multihop communication path connecting any two vertexes of the graph}. \qedsymbol

Due to the presence of the weighted norm in (\ref{Cent_MMSE_GMRF}) that couples the observations of all the nodes in the network,
the problem does not seem to be amenable for a distributed solution. However, since the precision matrix $\bB$ is a sparse matrix
that reflects the structure of the Markov graph, we have
\begin{equation}\label{potential}
V(\bx[k])=\frac{1}{2}\|\bx[k]-\bU[k]\boldsymbol{\theta}\|^2_{\boldsymbol{B}}=\sum_{i=1}^{N}V_i(\mathbf{x}_i[k]; \boldsymbol{\theta})
\end{equation}
with $\mathbf{x}_i[k] =[x_i[k], \{x_j[k]\}_{j \in {\cal M}_i, j>i}]^T$, and
\begin{align}\label{phi_functions}
V_i(\mathbf{x}_i[k];  \boldsymbol{\theta}):&=\frac{1}{2}\,b_{i,i}(x_i[k]-\bu^T_i[k]\boldsymbol{\theta})^2 \nonumber\\
&\hspace{-1.5cm}+\sum_{j\in{\cal M}_i, j>i}b_{i,j} (x_j[k]-\bu^T_j[k]\boldsymbol{\theta})(x_i[k]-\bu^T_i[k]\boldsymbol{\theta}),
\end{align}
where ${\cal M}_i=\{j\in V:b_{i,j}>0\}$ is the neighborhood of node $i$ in the statistical dependency graph. Thus, exploiting (\ref{potential}) in (\ref{Cent_MMSE_GMRF}), the global optimization problem becomes
\begin{equation}\label{Cent_MMSE_GMRF_2}
\boldsymbol{\hat{\theta}}[k]= \displaystyle {\rm arg} \min_{\boldsymbol{\theta}} \;\; \sum_{i=1}^{N}\mathbb{E}\left\{V_i(\mathbf{x}_i[k]; \boldsymbol{\theta})\right\}
\end{equation}
We follow the diffusion adaptation approach proposed in \cite{Cattivelli_Sayed, Chen-Sayed, SayedBC}, \cite{DiLorenzo-Sayed}, to devise
distributed strategies for the minimization of (\ref{Cent_MMSE_GMRF_2}). Thus, we introduce two sets of real, weighting coefficients $\bQ=\{q_{j,i}\}$, $\bW=\{w_{j,i}\}$ satisfying:
\begin{equation}\label{combination_coefficients}
q_{j,i}>0,\quad w_{j,i}>0  \quad \hbox{if} \quad j\in\mathcal{N}_i, \quad \bQ \mathbf{1}=\mathbf{1}, \quad \bW^T\mathbf{1}=\mathbf{1},
\end{equation}
where $\mathbf{1}$ denotes the $N\times1$ vector with unit entries and $\mathcal{N}_i$ denotes the spatial neighborhood of node $i$. Each coefficient $q_{j,i}$ (and $w_{j,i}$) represents a weight value that node $i$ assigns to information arriving from its neighbor $j$. Of course, the coefficient $q_{j,i}$ (and $w_{j,i}$) is equal to zero when nodes $j$ and $i$ are not directly connected. Furthermore, each row of $\bQ$ adds up to one so that the sum of all weights leaving each node $j$ should be one. Several diffusion strategies can then be derived from (\ref{Cent_MMSE_GMRF_2}), see e.g. \cite{Cattivelli_Sayed, Chen-Sayed, SayedBC}. For this purpose, we need to explicit the stochastic gradient of each potential function $V_i(\mathbf{x}_i[k];  \boldsymbol{\theta})$ in (\ref{Cent_MMSE_GMRF_2}), which can be written as:
\begin{align}\label{grad_phi}
&\nabla_{\boldsymbol{\theta}} V_i(\mathbf{x}_i[k];  \boldsymbol{\theta}) \;=\; -b_{i,i}\bu_i[k](x_i[k]-\bu_i^T[k]\boldsymbol{\theta}) \nonumber\\
\displaystyle&-\sum_{j\in\mathcal{M}_i;j>i} b_{i,j}\big[\big( x_j[k]\bu_i[k]+x_i[k]\bu_j[k]\big)-\big(\bu_j[k]\bu_i^T[k]\nonumber\\
&\hspace{1cm}+\bu_i[k]\bu_j^T[k]\big)\boldsymbol{\theta} \big]
\end{align}
The first algorithm that we consider is the Adapt-Then-Combine (ATC) strategy, which is reported in Table 1. We refer to this strategy as the ATC-GMRF diffusion LMS algorithm.
\begin{algorithm}
\caption*{\textbf{Table 1: ATC-GMRF diffusion LMS}}
\vspace{.3cm}
Start with $\boldsymbol{\theta}_{i}[-1]$ and $\boldsymbol{\psi}_{i}[-1]$ chosen at random for all $i$. Given non-negative real coefficients $\{q_{l,k},w_{l,k}\}$ satisfying (\ref{combination_coefficients2}), and sufficiently small step-sizes $\mu_i>0$, for each time $k\geq0$ and for each node $i$, repeat:
\begin{align}\label{ATC diffusion}
&\boldsymbol{\psi}_{i}[k]=\boldsymbol{\theta}_{i}[k-1]-\mu_i \displaystyle \sum_{j\in\mathcal{N}_i}q_{j,i}\nabla_{\boldsymbol{\theta}} V_j(\mathbf{x}_j[k];  \boldsymbol{\theta}_i[k-1]) \nonumber\\
&\hspace{4.3cm} \hbox{(adaptation step)} \\
&\boldsymbol{\theta}_{i}[k]=\displaystyle\sum_{j \in {\cal N}_i}w_{j,i}\boldsymbol{\psi}_{j}[k] \hspace{.9cm} \hbox{(combination step)}\nonumber
\end{align}
\end{algorithm}
The first step in Table 1 is an adaptation step, where the intermediate estimate $\boldsymbol{\psi}_{i}[k]$ is updated adopting
the stochastic gradients of the potential functions $V_j(\mathbf{x}_j[k];  \boldsymbol{\theta})$, $j\in\mathcal{N}_i$, in (\ref{phi_functions}).
As we can see from (\ref{grad_phi}), the evaluation of each gradient $\nabla_{\boldsymbol{\theta}} V_i(\mathbf{x}_i[k];  \boldsymbol{\theta}_i[k])$ requires
not only measurements from node $i$, but also data coming from nodes ${j \in {\cal M}_i}$, $j>i$, which are neighbors of $i$ in the dependency graph.
The coefficients $q_{j,i}$ determine which spatial neighbor nodes $j\in \mathcal{N}_i$ should share its measurements
with node $i$. The second step is a diffusion step where the intermediate
estimates $\boldsymbol{\psi}_{j}[k]$, from the spatial neighbors $j\in \mathcal{N}_i$, are combined through the weighting coefficients
$\{w_{j,i}\}$.
\begin{algorithm}
\caption*{\textbf{Table 2: CTA-GMRF diffusion LMS}}
\vspace{.3cm}
Start with $\boldsymbol{\theta}_{i}[-1]$ and $\boldsymbol{\chi}_{i}[-1]$ chosen at random for all $i$. Given non-negative real coefficients $\{q_{l,k},w_{l,k}\}$ satisfying (\ref{combination_coefficients2}), and sufficiently small step-sizes $\mu_i>0$, for each time $k\geq0$ and for each node $i$, repeat:
\begin{align}\label{CTA diffusion}
&\boldsymbol{\chi}_{i}[k-1]=\displaystyle\sum_{j \in {\cal N}_i}w_{j,i}\boldsymbol{\theta}_{j}[k-1] \hspace{.6cm} \hbox{(combination step)}\nonumber\\
&\hspace{5.2cm} \hbox{(adaptation step)} \\
&\boldsymbol{\theta}_{i}[k]=\boldsymbol{\chi}_{i}[k-1]-\mu_i \displaystyle \sum_{j\in\mathcal{N}_i}q_{j,i}\nabla_{\boldsymbol{\theta}} V_j(\mathbf{x}_j[k];  \boldsymbol{\chi}_i[k-1]) \nonumber
\end{align}
\end{algorithm}
We remark that a similar but alternative strategy, known as the Combine-then-Adapt (CTA) strategy, can also be derived, see, e.g.,
\cite{Cattivelli_Sayed, Chen-Sayed, SayedBC}; in this implementation, the only difference is that data aggregation is performed before
adaptation. We refer to this strategy as the CTA-GMRF diffusion LMS algorithm, and we report it in Table 2. The complexity of the GMRF
diffusion schemes in (\ref{ATC diffusion})-(\ref{CTA diffusion}) is $O(4M)$, i.e., they have linear complexity as standard stand-alone
LMS adaptation.

\noindent {\textbf{Remark 1:}}  As we can see from Tables 1 and 2 and eq. (\ref{grad_phi}), the GMRF diffusion LMS algorithms exploit
information coming from neighbors defined over two different graphs, i.e., the spatial adjacency graph and the statistical dependency graph.
In particular, the algorithms require that: i) each node exchanges information with its neighbors in the Markov dependency graph;
ii) the communication graph is connected in order to ensure that the data from an arbitrary sensor can percolate through the entire
network. These conditions must be guaranteed by a proper design of the communication graph, which should contain the Markov dependency graph
as a subgraph. This represents a generalization of the distributed computation observed in the conditionally independent case, where the exchange
of information among nodes takes into account only the spatial proximity of nodes \cite{Cattivelli_Sayed}. In the more general Markovian case,
the organization of the communication network should take into account, {\it jointly}, the grouping suggested by the cliques of the
underlying dependency graph.

\section{Mean-Square Performance Analysis}

From now on, we view the estimates $\boldsymbol{\theta}_{i}[k]$ as realizations of a random process and analyze the performance of the diffusion algorithm over GMRF in terms of its mean-square behavior. Following similar arguments as in \cite{Cattivelli_Sayed}, we formulate a general form that includes the ATC and CTA algorithms as special cases. Thus, consider a general diffusion filter of the form
\begin{align}\label{ATCTA diffusion}
&\boldsymbol{\chi}_{i}[k-1]=\displaystyle\sum_{j \in {\cal N}_i}p^{(1)}_{j,i}\boldsymbol{\theta}_{j}[k-1]\nonumber\\
&\boldsymbol{\psi}_{i}[k]=\boldsymbol{\chi}_{i}[k-1]-\mu_i \displaystyle \sum_{j\in\mathcal{N}_i}s_{j,i}\nabla_{\boldsymbol{\theta}} V_j(\mathbf{x}_j[k];  \boldsymbol{\chi}_i[k-1]) \\
&\boldsymbol{\theta}_{i}[k]=\displaystyle\sum_{j \in {\cal N}_i}p^{(2)}_{j,i}\boldsymbol{\psi}_{j}[k] \nonumber
\end{align}
where the coefficients $p^{(1)}_{j,i}$, $s_{j,i}$, and $p^{(2)}_{j,i}$ are generic non-negative real coefficients corresponding to the entries of matrices $\bP_{1}$, $\bS$ and $\bP_{2}$, respectively, and satisfy
\begin{equation}\label{combination_coefficients2}
\bP_{1}^T\mathbf{1}=\mathbf{1}, \quad \bS \mathbf{1}=\mathbf{1}, \quad \bP_{2}^T\mathbf{1}=\mathbf{1}.
\end{equation}
Equation (\ref{ATCTA diffusion}) can be specialized to the ATC-GMRF diffusion LMS algorithm (\ref{ATC diffusion}) by choosing $\bP_{1}=\bI$, $\bS=\bQ$ and $\bP_{2}=\bW$, and to the CTA-GMRF diffusion LMS algorithm (\ref{CTA diffusion}) by choosing $\bP_{1}=\bW$, $\bS=\bQ$ and $\bP_{2}=\bI$. To proceed with the analysis, we introduce the error quantities $\tilde{\boldsymbol{\theta}}_{i}[k]=\boldsymbol{\theta}_0-\boldsymbol{\theta}_{i}[k]$, $\tilde{\boldsymbol{\chi}}_{i}[k]=\boldsymbol{\theta}_0-\boldsymbol{\chi}_{i}[k]$, $\tilde{\boldsymbol{\psi}}_{i}[k]=\boldsymbol{\theta}_0-\boldsymbol{\psi}_{i}[k]$, and the network vectors:
\begin{eqnarray}\label{err_vectors}
\tilde{\boldsymbol{\theta}}[k]=\begin{bmatrix} \tilde{\boldsymbol{\theta}}_{1}[k] \\ \vdots \\ \tilde{\boldsymbol{\theta}}_{N}[k]  \end{bmatrix},\hspace{.1cm}
\tilde{\boldsymbol{\chi}}[k]=\begin{bmatrix} \tilde{\boldsymbol{\chi}}_{1}[k] \\ \vdots \\ \tilde{\boldsymbol{\chi}}_{N}[k]  \end{bmatrix},\hspace{.1cm}
\tilde{\boldsymbol{\psi}}[k]=\begin{bmatrix} \tilde{\boldsymbol{\psi}}_{1}[k] \\ \vdots \\ \tilde{\boldsymbol{\psi}}_{N}[k]  \end{bmatrix}
\end{eqnarray}
We also introduce the block diagonal matrix
\begin{eqnarray}\label{step_matrix}
\bM={\rm diag}\{\mu_1\bI_{M},\ldots,\mu_N\bI_{M}\}
\end{eqnarray}
and the extended block weighting matrices
\begin{eqnarray}\label{combination_matrices}
\hat{\bP}_1=\bP_1\otimes \bI_{M}, \hspace{.3cm} \hat{\bS}=\bS\otimes \bI_{M}, \hspace{.3cm} \hat{\bP}_2=\bP_2\otimes \bI_{M}
\end{eqnarray}
where $\otimes$ denotes the Kronecker product operation. We further introduce the random block quantities:
\begin{align}
&\bD[k]\;=\;{\rm diag}\bigg\{\sum_{j\in{\cal N}_i}s_{j,i}\bigg[b_{j,j}\bu_j[k]\bu_j^T[k] \nonumber\\
&\hspace{.5cm}+\sum_{l\in{\cal M}_j; l>j}b_{j,l}\left(\bu_l[k]\bu_j^T[k]+\bu_j[k]\bu_l^T[k]\right)\bigg]\bigg\}_{i=1}^N \label{perf_matrices}\\
&\bg[k]\;=\;\hat{\bS}^T\cdot{\rm col}\bigg\{b_{i,i}\bu_{i}[k]\bv_i[k]\label{perf_matrices2}\\
&\hspace{.3cm}+\sum_{j\in{\cal M}_i; j>i}b_{i,j}\big(\bu_i[k]v_j[k]+\bu_j[k]v_i[k]\big)\bigg\}_{i=1}^N
=\hat{\bS}^T \hat{\bg}[k] \nonumber
\end{align}
Then, exploiting the linear observation model in (\ref{linear_model}), we conclude from (\ref{grad_phi})-(\ref{ATCTA diffusion}) that the following relations hold for the error vectors:
\begin{align}
\tilde{\boldsymbol{\chi}}[k-1]&=\hat{\bP}_1^T \tilde{\boldsymbol{\theta}}[k-1]\nonumber\\
\tilde{\boldsymbol{\psi}}[k]&=\tilde{\boldsymbol{\chi}}[k-1]-\bM\left(\bD[k]\tilde{\boldsymbol{\chi}}[k-1]+\bg[k]\right)\label{psi}\\
\tilde{\boldsymbol{\theta}}[k]&=\hat{\bP}_2^T\tilde{\boldsymbol{\psi}}[k] \nonumber
\end{align}
We can combine the equations in (\ref{psi}) into a single recursion:
\begin{eqnarray}\label{compact_Diffusion}
\tilde{\boldsymbol{\theta}}[k]=\hat{\bP}_2^T\left(I-\bM\bD[k]\right)\hat{\bP}_1^T\tilde{\boldsymbol{\theta}}[k-1]-\hat{\bP}_2^T\bM\bg[k]
\end{eqnarray}
This relation tells us how the network weight-error vector evolves over time. The relation will be the launching point for our mean-square analysis. To proceed, we introduce the following independence assumption on the regression data.

\noindent {\textbf{Assumption 2 (Independent regressors)}} {\it The regressors $\bu_i[k]$ are temporally white and spatially independent with $\bR_{u,i}=\mathbb{E}\{\bu_{i}[k]\bu^T_{i}[k]\}\succ0$.}{\qedsymbol}

It follows from Assumption 2 that $\bu_{i}[k]$ is independent of $\{\boldsymbol{\theta}_j[t]\}$ for all $j$ and $t\leq k-1$. Although not true in general, this assumption is common in the adaptive filtering literature since it helps simplify the analysis. Several studies in the literature, especially on stochastic approximation theory \cite{Kushner-Yin}--\cite{Sayed}, indicate that the performance expressions obtained using this assumption match well the actual performance of stand-alone filters for sufficiently small step-sizes. Therefore, we shall also rely on the following condition.

\noindent {\textbf{Assumption 3 (Small step-sizes)}} {\it The step-sizes $\{\mu_i\}$ are sufficiently small so that terms that depend on higher-order powers of $\mu_i$ can be ignored.}{\qedsymbol}

\subsection{Convergence in the Mean}

Exploiting eq. (\ref{perf_matrices}) and Assumption 2, we have
\begin{eqnarray}\label{perf_matrices3}
\bD\triangleq\mathbb{E}\{\bD[k]\}={\rm diag}\left\{\sum_{j\in{\cal N}_i}s_{j,i}b_{j,j}\bR_{u,j},\right\}_{i=1}^N
\end{eqnarray}
Then, taking expectations of both sides of (\ref{compact_Diffusion}) and calling upon Assumption 1, we conclude that the mean-error vector evolves according to the following dynamics:
\begin{eqnarray}\label{compact_Diffusion_Expected}
\mathbb{E}\tilde{\boldsymbol{\theta}}[k]=\hat{\bP}_2^T\left(I-\bM\bD\right)\hat{\bP}_1^T\mathbb{E}\tilde{\boldsymbol{\theta}}[k-1]
\end{eqnarray}
The following theorem guarantees the asymptotic mean stability of the diffusion strategies over GMRF (\ref{ATC diffusion})-(\ref{CTA diffusion}).

\noindent {\textbf{Theorem 1 (Stability in the mean)}} \textit{ Assume data model (\ref{linear_model}) and Assumption 2 hold. Then, for any initial condition and any choice of the matrices $\bQ$ and $\bW$ satisfying (\ref{combination_coefficients}), the diffusion strategies (\ref{ATC diffusion})-(\ref{CTA diffusion}) asymptotically converges in the mean if the step-sizes are chosen to satisfy:
\begin{eqnarray}\label{step_sizes}
0<\mu_i<\frac{2}{\lambda_{\max}\left\{\sum_{j\in{\cal N}_i}s_{j,i}b_{j,j}\bR_{u,j}\right\}} \quad i=1,\ldots,N,
\end{eqnarray}
where $\lambda_{\max}(\bX)$ denotes the maximum eigenvalue of a Hermitian positive semi-definite matrix $\bX$.}

\begin{proof}
See Appendix A.
\end{proof}

\subsection{Convergence in Mean-Square}

We now examine the behavior of the steady-state mean-square deviation, $\mathbb{E}\|\tilde{\boldsymbol{\theta}}_i[k]\|^2$ as $k\rightarrow\infty$. Following the energy conservation framework of \cite{Lopes_Sayed,Cattivelli_Sayed} and under Assumption 2, from (\ref{compact_Diffusion}), we can establish the following variance relation:
\begin{eqnarray}\label{weighted_norm_expanded}
\mathbb{E}\|\tilde{\boldsymbol{\theta}}[k]\|^2_{\boldsymbol{\Sigma}}=\mathbb{E}\|\tilde{\boldsymbol{\theta}}[k-1]\|^2_{\boldsymbol{\Sigma}'}+ \mathbb{E}[\bg^T[k]\bM\hat{\bP}_2\boldsymbol{\Sigma}\hat{\bP}_2^T\bM\bg[k]]
\end{eqnarray}
where $\boldsymbol{\Sigma}$ is any Hermitian nonnegative-definite matrix that we are free to choose, and
\begin{eqnarray}\label{Sigma'}
\boldsymbol{\Sigma}'=\hat{\bP}_1(\bI-\bD[k]\bM)\hat{\bP}_2\boldsymbol{\Sigma} \hat{\bP}_2^T(\bI-\bM\bD[k])\hat{\bP}_1^T \label{Sigma_primo}
\end{eqnarray}
Now, from eq. (\ref{perf_matrices2}), let us define
\begin{eqnarray}\label{G_matrix}
\bG=\mathbb{E}[\bg[k]\bg^T[k]]=\bS^T\mathbb{E}[\hat{\bg}[k]\hat{\bg}^T[k]]\bS=\bS^T\hat{\bG}\bS
\end{eqnarray}
where $\hat{\bG}=\mathbb{E}[\hat{\bg}[k]\hat{\bg}^T[k]]$ is an $MN\times MN$ block matrix, where each block $\hat{\bG}_{i,j}$ is an $M\times M$ matrix. Exploiting Assumption 2 and eq. (\ref{perf_matrices2}), the $(i,i)$-th block of matrix $\hat{\bG}$, $i=1,\ldots,N$, is given by
\begin{align}\label{Gii}
\hat{\bG}_{i,i}&=\mathbb{E}[\bg_i[k]\bg_i^T[k]]= \bR_{u,i}\bigg[c_{i,i}b_{i,i}^2+2b_{i,i}\sum_{j\in{\cal M}_i; j>i}b_{i,j}c_{i,j}\nonumber\\
&\hspace{-.5cm}+\mathbb{E}\bigg(\sum_{j\in{\cal M}_i; j>i}b_{i,j}v_j[k]\bigg)^2\bigg]+c_{i,i}\sum_{j\in{\cal M}_i; j>i}b^2_{i,j}\bR_{u,j} \end{align}
where the third term in term in (\ref{Gii}) can be expressed in closed form. Indeed, defining the set $\mathcal{A}_i=\{j\in{\cal M}_i, j>i\}$ and associating each term $b_{i,j}v_j[k]$, $j\in\Omega_i$, to the term $x_t$, $t=1,\ldots,m$, $m=\textrm{card}\{\mathcal{A}_i\}$, from a direct application of the Multinomial theorem \cite{Graham}, we have
\begin{align}\label{Multinomial}
&\mathbb{E}\left(\sum_{j\in{\cal M}_i; j>i}b_{i,j}v_j[k]\right)^2=\mathbb{E}\left(\sum_{t=1}^{m} x_t\right)^2\nonumber\\
&=\sum_{h_1+\ldots+h_m=2}\left(
                           \begin{array}{c}
                             2 \\
                             h_1\ldots h_m\\
                           \end{array}
                         \right) \mathbb{E}\left\{x_1^{h_1}\cdots x_m^{h_m}\right\}
\end{align}
where the products in (\ref{Multinomial}) have only quadratic terms such that
\begin{align}
&\mathbb{E}\left\{x_t^2\right\}\rightarrow \mathbb{E}\left\{b^2_{i,j}v^2_j[k]\right\}=b^2_{i,j}c_{j,j} \\
&\mathbb{E}\left\{x_tx_s\right\}\rightarrow \mathbb{E}\left\{b_{i,j}b_{i,l}v_j[k]v_l[k]\right\}=b_{i,j}b_{i,l}c_{j,l}
\end{align}
with $t,s=1,\ldots,m$, and $j,l\in\mathcal{A}_i$. At the same way, the $(i,l)$-th block of matrix $\hat{\bG}$, $i,l=1,\ldots,N$, is given by
 \begin{align}\label{Gij}
&\hat{\bG}_{i,l}=\mathbb{E}[\hat{\bg}_i[k]\hat{\bg}_l^T[k]] \;=\; c_{i,l}\sum_{n\in \mathcal{A}_i\cap\mathcal{A}_l} b_{i,n}b_{l,n}\bR_{u,n} \nonumber\\
&+\bR_{u,i} b_{i,l}\left[b_{i,i}c_{i,l}+\left(\sum_{j\in{\cal M}_i; j>i}b_{i,j}c_{j,l}\right)\right] I(i\in\mathcal{A}_l)   \nonumber\\
&\hspace{-.28cm}+\bR_{u,l} b_{i,l}\left[b_{l,l}c_{i,l}+\left(\sum_{m\in{\cal M}_l; m>l}b_{l,m}c_{i,m}\right)\right] I(l\in\mathcal{A}_i)
\end{align}
where $I(\mathcal{Y})$ is the indicator function, i.e. $I(\mathcal{Y})=1$ if the event $\mathcal{Y}$ is true and zero otherwise, and $\mathcal{A}_l=\{m\in{\cal M}_l; m>l\}$. Then, given the closed form expression for the matrix $\bG$ given by eqs. (\ref{G_matrix})-(\ref{Gij}), we can rewrite recursion (\ref{weighted_norm_expanded}) as:
\begin{eqnarray}\label{weighted_norm2}
\mathbb{E}\|\tilde{\boldsymbol{\theta}}[k]\|^2_{\boldsymbol{\Sigma}}=\mathbb{E}\|\tilde{\boldsymbol{\theta}}[k-1]\|^2_{\boldsymbol{\Sigma}'}+{\rm Tr}[\boldsymbol{\Sigma} \hat{\bP}_2^T\bM\bG\bM\hat{\bP}_2]
\end{eqnarray}
where ${\rm Tr}(\cdot)$ denotes the trace operator. Let $\boldsymbol{\sigma}={\rm vec}(\boldsymbol{\Sigma})$ and $\boldsymbol{\sigma}'=\mbox{\rm vec}(\boldsymbol{\Sigma}')$,
where the ${\rm vec}(\cdot)$ notation stacks the columns of $\boldsymbol{\Sigma}$ on top of each other and ${\rm vec}^{-1}(\cdot)$ is the inverse operation.
Using the Kronecker product property ${\rm vec}(\bU\boldsymbol{\theta} \bV)=(\bV^T\otimes \bU){\rm vec}(\boldsymbol{\Sigma})$,
we can vectorize both sides of (\ref{Sigma'}) and conclude that (\ref{Sigma'}) can be replaced by the simpler linear vector relation: $\boldsymbol{\sigma}'={\rm vec}(\boldsymbol{\Sigma}')=\bF\boldsymbol{\sigma}$,
where $\bF$ is the following $N^2M^2\times N^2M^2$ matrix with block entries of size $M^2\times M^2$ each:
\begin{align}\label{matrix_F}
\bF\;=\;&(\hat{\bP}_1\otimes \hat{\bP}_1)\big\{\bI-\bI\otimes(\bD\bM)-(\bD^T\bM)\otimes \bI \nonumber\\ &\;+\mathbb{E}(\bD^T[k]\bM)\otimes(\bD[k]\bM)\big\}(\hat{\bP}_2\otimes \hat{\bP}_2)
\end{align}
Using the property ${\rm Tr}(\boldsymbol{\Sigma} \bX)={\rm vec}(\bX^T)^T\boldsymbol{\sigma}$
we can then rewrite (\ref{weighted_norm2}) as follows:
\begin{align}\label{weighted_norm3}
\mathbb{E}\|\tilde{\boldsymbol{\theta}}[k]\|^2_{\textrm{vec}^{-1}(\boldsymbol{\sigma})}\;=\;&\mathbb{E}\|\tilde{\boldsymbol{\theta}}[k-1]\|^2_{\textrm{vec}^{-1}(\boldsymbol{F\sigma})}\nonumber\\
&+[{\rm vec}(\hat{\bP}_2^T\bM\bG^T\bM\hat{\bP}_2)]^T\boldsymbol{\sigma}
\end{align}
The following theorem guarantees the asymptotic mean-square stability (i.e., convergence in the mean and mean-square sense) of the diffusion strategies over GMRF in (\ref{ATC diffusion})-(\ref{CTA diffusion}).

\noindent {\textbf{Theorem 2 (Mean-Square Stability)}} \textit{Assume model (\ref{linear_model}) and Assumption 1 hold. Then, the GMRF diffusion LMS algorithms (\ref{ATC diffusion})-(\ref{CTA diffusion}) will be mean-square stable if the step-sizes are such that (\ref{step_sizes}) is satisfied and the matrix $\bF$ in (\ref{matrix_F}) is stable.}

\begin{proof}
See Appendix B.
\end{proof}

\noindent {\textbf{Remark 2:}} Note that the step sizes influence (\ref{matrix_F}) through the matrix $\bM$.
Since in virtue of Assumption 2 the step-sizes are sufficiently small, we can ignore terms that depend on higher-order powers of the step-sizes. Then, we approximate (\ref{matrix_F}) as
\begin{align}\label{approx_matrix_F}
\bF \;\approx\; &(\hat{\bP}_1\otimes \hat{\bP}_1)\big\{\bI-\bI\otimes (\bD \bM)-(\bD^T\bM)\otimes \bI \nonumber\\
&+(\bD^T\bM)\otimes (\bD\bM)\big\}(\hat{\bP}_2\otimes \hat{\bP}_2)=\bH^T\otimes\bH
\end{align}
where $\bH=\hat{\bP}_2^T(\bI-\bM\bD)\hat{\bP}_1^T$. Now, since $\hat{\bP}_1$ and $\hat{\bP}_2$ are left-stochastic, it can be verified that the above $\bF$ is stable if $\bI-\bD\bM$ is stable \cite{Chen-Sayed, SayedBC}; this latter condition is guaranteed by (\ref{step_sizes}).
In summary, sufficiently small step-sizes ensure the stability of the diffusion strategies over GMRF in the mean and mean-square senses. {\qedsymbol}

\subsection{Mean-Square Performance}

Taking the limit as $k \rightarrow \infty$ (assuming the step-sizes are small enough to ensure convergence to a steady-state), we deduce from (\ref{weighted_norm3}) that:
\begin{eqnarray}\label{variance_relation2}
\displaystyle \lim_{k\rightarrow\infty}\mathbb{E}\|\tilde{\boldsymbol{\theta}}[k]\|^2_{\textrm{vec}^{-1}\left((\boldsymbol{I}-\boldsymbol{F})\boldsymbol{\sigma}\right)}=[{\rm vec}(\hat{\bP}_2^T\bM\bG^T\bM\hat{\bP}_2)]^T\boldsymbol{\sigma}
\end{eqnarray}
Expression (\ref{variance_relation2}) is a useful result: it allows us to derive several performance metrics through the proper
selection of the free weighting parameter $\boldsymbol{\sigma}$ (or $\boldsymbol{\Sigma}$), as was done in \cite{Cattivelli_Sayed}.
For example, the MSD for any node $k$ is defined as the steady-state value $\mathbb{E}\|\tilde{\boldsymbol{\theta}}_i[k]\|^2$, as
$k\rightarrow\infty$, and can be obtained by computing $\lim_{k\rightarrow\infty}\mathbb{E}\|\tilde{\boldsymbol{\theta}}[k]\|^2_{\boldsymbol{T}_i}$ with a block weighting matrix $\bT_i$
that has the $M\times M$ identity matrix at block $(i,i)$ and zeros elsewhere. Then, denoting the vectorized version of the matrix $\bT_i$ by
$\bt_i={\rm vec}({\rm diag}(\bee_i)\otimes \bI_M)$, where $\bee_i$ is the vector whose $i$-th entry is one and zeros elsewhere,
and if we select $\boldsymbol{\sigma}$ in (\ref{variance_relation2}) as $\boldsymbol{\sigma}_i=(\bI-\bF)^{-1}\bt_i$, we arrive at the
MSD for node $i$:
\begin{eqnarray}\label{MSD_k}
{\rm MSD}_i=[{\rm vec}(\hat{\bP}_2^T\bM\bG^T\bM\hat{\bP}_2)]^T(\bI-\bF)^{-1}\bt_i
\end{eqnarray}
The average network ${\rm MSD}_{\rm net}$ is given by:
\begin{eqnarray}\label{MSD}
{\rm MSD}_{\rm net}=\displaystyle \lim_{k\rightarrow\infty}\frac{1}{N}\sum_{i=1}^N\mathbb{E}\|\tilde{\boldsymbol{\theta}}_i[k]\|^2
\end{eqnarray}
Then, to obtain the network MSD from (\ref{variance_relation2}), the weighting matrix of
$\lim_{k\rightarrow\infty}\mathbb{E}\|\tilde{\boldsymbol{\theta}}[k]\|^2_{\boldsymbol{T}}$ should be chosen as $\bT=\bI_{MN}/N$.
Let $\bt$ denote the vectorized version of $\bI_{MN}$, i.e., $\bt={\rm vec}(\bI_{MN})$, and selecting $\boldsymbol{\sigma}$ in
(\ref{variance_relation2}) as $\boldsymbol{\sigma}=(\bI-\bF)^{-1}\bt/N$, the network MSD is given by:
\begin{eqnarray}\label{MSD_net}
{\rm MSD}_{net}=\frac{1}{N}[{\rm vec}(\hat{\bP}_2^T\bM\bG^T\bM\hat{\bP}_2)]^T(\bI-\bF)^{-1}\bt
\end{eqnarray}
In the sequel, we will confirm the validity of these theoretical expressions by comparing them with numerical results.

\section{Sparse Diffusion Adaptation over Gaussian Markov Random Fields}

In this section, we extend the previous algorithms by incorporating thresholding functions that can help improving the performance of the diffusion LMS algorithm over GMRF under a sparsity assumption of the vector $\boldsymbol{\theta}_0$ to be estimated. Since it was argued in \cite{Cattivelli_Sayed} that ATC strategies generally outperform CTA strategies, we continue our discussion by focusing on extensions of the ATC algorithm (\ref{ATC diffusion}); similar arguments applies to CTA strategies. The main idea is to add a sparsification step in the processing chain of the ATC strategy (\ref{ATC diffusion}), in order to drive the algorithm toward a sparse estimate. In this paper, we consider two main strategies. The first strategy performs the sparsification step after the adaptation and combination steps. We will refer to this strategy as the ACS-GMRF diffusion LMS algorithm, and its main steps are reported in Table 3.
\begin{algorithm}
\caption*{\textbf{Table 3: ACS-GMRF diffusion LMS}}
\vspace{.3cm}
Start with $\boldsymbol{\theta}_{i}[-1]$, $\boldsymbol{\psi}_{i}[-1]$, $\boldsymbol{\zeta}_{i}[-1]$ chosen at random for all $i$. Given non-negative real coefficients $\{q_{l,k},w_{l,k}\}$ satisfying (\ref{combination_coefficients2}), and sufficiently small step-sizes $\mu_i>0$, for each time $k\geq0$ and for each node $i$, repeat:
\begin{align}\label{ACS diffusion}
&\boldsymbol{\psi}_{i}[k]=\boldsymbol{\theta}_{i}[k-1]-\mu_i \displaystyle \sum_{j\in\mathcal{N}_i}q_{j,i}\nabla_{\boldsymbol{\theta}} V_j(\mathbf{x}_j[k];  \boldsymbol{\theta}_i[k-1]) \nonumber\\
&\hspace{4.1cm} \hbox{(adaptation step)} \\
&\boldsymbol{\zeta}_{i}[k]=\displaystyle\sum_{j \in {\cal N}_i}w_{j,i}\boldsymbol{\psi}_{j}[k] \hspace{.65cm} \hbox{(combination step)}\nonumber\\
&\boldsymbol{\theta}_{i}[k]=\bT_\gamma\left(\boldsymbol{\zeta}_{i}[k]\right) \hspace{1.2cm} \hbox{(sparsification step)}\nonumber
\end{align}
\end{algorithm}
The second strategy performs instead the sparsification step in the middle between adaptation and combination steps, as we can notice from Table 4. We will refer to it as the ASC-GMRF diffusion LMS algorithm.
\begin{algorithm}
\caption*{\textbf{Table 4: ASC-GMRF diffusion LMS}}
\vspace{.3cm}
Start with $\boldsymbol{\theta}_{i}[-1]$, $\boldsymbol{\psi}_{i}[-1]$, $\boldsymbol{\zeta}_{i}[-1]$ chosen at random for all $i$. Given non-negative real coefficients $\{q_{l,k},w_{l,k}\}$ satisfying (\ref{combination_coefficients2}), and sufficiently small step-sizes $\mu_i>0$, for each time $k\geq0$ and for each node $i$, repeat:
\begin{align}\label{ASC diffusion}
&\boldsymbol{\psi}_{i}[k]=\boldsymbol{\theta}_{i}[k-1]-\mu_i \displaystyle \sum_{j\in\mathcal{N}_i}q_{j,i}\nabla_{\boldsymbol{\theta}} \phi_j(\mathbf{x}_j[k];  \boldsymbol{\theta}_i[k-1]) \nonumber\\
&\hspace{4.1cm} \hbox{(adaptation step)} \\
&\boldsymbol{\zeta}_{i}[k]=\bT_\gamma\left(\boldsymbol{\psi}_{i}[k]\right) \hspace{1.2cm} \hbox{(sparsification step)}\nonumber\\
&\boldsymbol{\theta}_{i}[k]=\displaystyle\sum_{j \in {\cal N}_i}w_{j,i}\boldsymbol{\zeta}_{j}[k] \hspace{.8cm} \hbox{(combination step)}\nonumber
\end{align}
\end{algorithm}
The sparsifcation step in Tables 3 and 4 is performed by using a thresholding function $\bT_\gamma(\bx)$. Several different functions can be used to enforce sparsity. A commonly used thresholding function comes directly by imposing an $\ell_1$ norm constraint in (\ref{Cent_MMSE_GMRF}), which is commonly known as the LASSO \cite{Tibshirani}. In this case, the vector threshold function $\bT_\gamma(\bx)$ is the component-wise thresholding function $T_\gamma(x_m)$ applied to each element $x_m$ of vector $\bx$, with
      \begin{align} \label{Lasso}
         T_\gamma(x_m)=\left\{
                      \begin{array}{ll}
                        x_m-\gamma, & \hbox{$x_m>\gamma$;} \\
                        0, & \hbox{$-\gamma\leq x_m \leq \gamma$;} \\
                        x_m+\gamma, & \hbox{$x_m<-\gamma$.}
                      \end{array}
                    \right.
      \end{align}
$m=1,\ldots,M$. The function $\bT_\gamma(\bx)$ in (\ref{Lasso}) tends to shrink all the components of the vector $\bx$ and, in particular, attracts to zero the components whose magnitude is within the threshold $\gamma$. We denote the strategy using this function as the $\ell_1$-ACS-GMRF diffusion LMS algorithm (or its ASC version). Since the LASSO constraint is known for introducing a bias in the estimate, the performance would deteriorate for vectors that are not sufficiently sparse. To reduce the bias introduced by the LASSO constraint, several other thresholding functions can be adopted to improve the performance also in the case of less sparse systems. A potential improvement can be made by modifying the thresholding function $\bT_\gamma(\bx)$ in (\ref{Lasso}) as
      \begin{align} \label{Rw_l1}
                    \hspace{-.3cm} T_\gamma(x_m)=\left\{
                      \begin{array}{ll}
                        \hspace{-.15cm}x_m-\gamma \;{\rm sign}(x_m), \hspace{-.1cm}& \hbox{$\displaystyle |x_m|>\gamma f(\varepsilon+|x_m|)$;} \\\\
                        0, & \hspace{-.9cm} \hbox{elsewhere;}
                      \end{array}
                    \right.
      \end{align}
$m=1,\ldots,M$, where $0<\varepsilon \ll 1$ denotes a small positive weight, $f(y)=1/y$, for $y\leq1$, and $f(y)=1$ elsewhere.
Compared to (\ref{Lasso}), the function in (\ref{Rw_l1}) adapts the threshold $\gamma\cdot f(\varepsilon+|x_m|)$ according to
the magnitude of the components \cite{Candes-Wakin-Boyd}. When the components are small with respect to $\varepsilon$,
the function in (\ref{Rw_l1}) increases its threshold so that the components are attracted to zero with a larger probability,
whereas, in the case of large components, the threshold is increased to ensure a small effect on them. We denote the strategy
using the function in (\ref{Rw_l1}) as the reweighted-$\ell_1$-ACS-GMRF diffusion LMS algorithm (or its ASC version).
The reweighted $\ell_1$ estimator in (\ref{Rw_l1}) is supposed to give better performance than the LASSO. Nevertheless,
it still might induce a too large bias if the vector is not sufficiently sparse. To further reduce the effect of the bias,
we consider the non-negative GAROTTE estimator as in \cite{Yuan-Lin}, whose thresholding function is defined as a vector whose
entries are derived applying the threshold
      \begin{align} \label{Garotte}
                    T_\gamma(x_m)=\left\{
                      \begin{array}{ll}
                        x_m\;(1-\gamma^2/x_m^2), \hspace{.5cm}& \hbox{$|x_m|>\gamma$;} \\\\
                        0, & \hspace{-.9cm} \hbox{$-\gamma\leq x_m \leq \gamma$;}
                      \end{array}
                    \right.
      \end{align}
$m=1,\ldots,M$. We denote the strategy using the function in (\ref{Garotte}) as the G-ACS-GMRF diffusion LMS algorithm (or its ASC version).
Ideally, sparsity is better represented by the $\ell_0$ norm as the regularization factor in (\ref{Cent_MMSE_GMRF}); this norm denotes the
number of non-zero entries in a vector. Considering that $\ell_0$ norm minimization is an NP-hard problem, the $\ell_0$ norm
is generally approximated by a continuous function. A popular approximation \cite{Gu-Jin-Mei,Liu-Li-Zhang} is
\begin{equation}\label{l0_norm}
\|\bx\|_0\simeq\sum_{m=1}^{M}\left(1-e^{-\beta\left|x_m\right|}\right),
\end{equation}
where $\beta>0$ is a shape parameter.
Based on a first order Taylor approximation of (\ref{l0_norm}), the thresholding function associated to the $\ell_0$ norm can be expressed as \cite{Liu-Liu-Li}:
\begin{align} \label{l0}
   T_\gamma(x_m)=\left\{
        \begin{array}{ll}
         x_m, & \hbox{$|x_m|>1/\beta$;} \\
        \frac{x_m-\beta\gamma\cdot{\rm sign}(x_m)}{1-\gamma\beta^2}, & \hbox{$\gamma\beta<|x_m|<1/\beta$;} \\
         0, & \hbox{$|x_m|<\gamma\beta$}
        \end{array}
          \right.
\end{align}
$m=1,\ldots,M$, with $\beta<\sqrt{1/\gamma}$. We can see how the $\ell_0$ thresholding function takes non-uniform effects on different components, and shrinks the small components around zero. We denote the strategy using the function in (\ref{l0}) as the $\ell_0$-ACS-GMRF diffusion LMS algorithm (or its ASC version). In the sequel, numerical results will show the performance achieved by adopting the thresholding functions in (\ref{Lasso}), (\ref{Rw_l1}), (\ref{Garotte}), and (\ref{l0}).

\noindent {\textbf{Remark 3:}} It is important to highlight the pros and cons of the proposed strategies in (\ref{ACS diffusion}) and
(\ref{ASC diffusion}). The adoption of the thresholding functions in (\ref{Lasso})-(\ref{l0}), determines that, if the vector
$\boldsymbol{\theta}_0$ is sparse, after the sparsification step only a subset of the entries of the local estimates are different from zero.
 Indeed, this thresholding operation allows to estimate the support of the vector $\boldsymbol{\theta}_0$, i.e., the set of indices of the non-zero
  component, which is denoted by ${\rm supp}(\boldsymbol{\theta}_0) = \{m : \theta_{0,m} \neq 0\}$.
Now, since in the ACS strategy in (\ref{ACS diffusion}) the combination step is performed before the sparsification, the thresholding function
will be able to correctly identify the zero entries of the vector with larger probability with respect to the ASC strategy in
(\ref{ASC diffusion}), thanks to the noise reduction effect due to the cooperation among nodes. At the same time,
sparsifying the vector before the combination step, as it is performed in the ASC strategy, has the advantage that, if the vector is
very sparse, each node must transmit to its neighbors only the few entries belonging to the estimated vector support, thus remarkably
reducing the burden of information exchange. This intuition suggests that the two strategies lead to an interesting tradeoff between
 performance and communication burden, as we will illustrate in the numerical results. \qedsymbol

The following theorem guarantees the asymptotic mean-square stability (i.e., stability in the mean and mean-square sense) of the sparse diffusion strategies over GMRF in (\ref{ACS diffusion})-(\ref{ASC diffusion}). Interestingly, stability is guaranteed under the same conditions of the sparsity agnostic strategies in (\ref{ATC diffusion})-(\ref{CTA diffusion}).

\noindent {\textbf{Theorem 3 (Mean-Square Stability)}} \textit{Assume model (\ref{linear_model}) and Assumption 2 hold. Then, the sparse diffusion strategies over GMRF (\ref{ACS diffusion})-(\ref{ASC diffusion}) will be mean-square stable if condition (\ref{step_sizes}) is satisfied and the matrix $\bF$ in (\ref{matrix_F}) is stable.}

\begin{proof}
See Appendix C.
\end{proof}

\section{Numerical Results}

In this section, we provide some numerical examples to illustrate the performance of the diffusion strategies over GMRF. In the first example,
we evaluate the performance of the proposed strategies, comparing it with respect to standard diffusion algorithms from \cite{Cattivelli_Sayed}.
The second example shows the benefits of using the ACS and ASC strategies in (\ref{ACS diffusion})-(\ref{ASC diffusion}) in the case
of sparseness of the vector to be estimated. The third example illustrates the capability of the proposed strategies to track time-varying,
sparse vector parameters.

\begin{figure}[t]
\includegraphics[width=8.3cm]{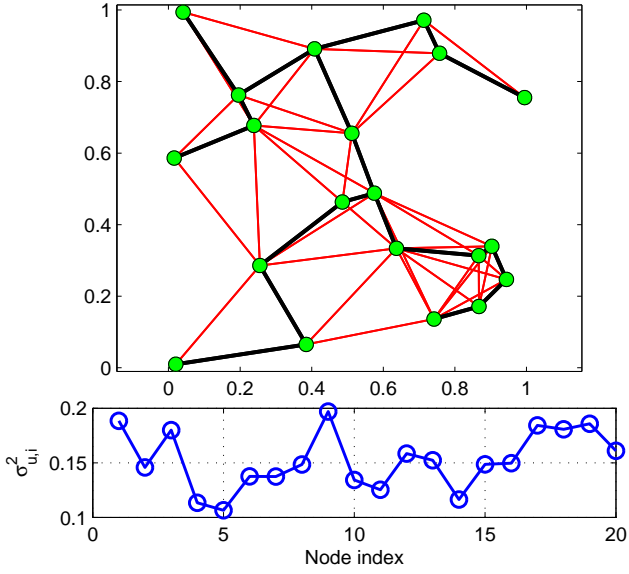}
  \caption{Adjacency (all the links) and Dependency (thick links) graphs (top), and regressor variances (bottom).}\label{net}
\end{figure}

\vspace{.2cm}
{\it Numerical Example - Performance :} We consider a connected network composed of 20 nodes. The spatial topology of the network
is depicted in Fig. \ref{net} (all the links are communication links). The regressors $\bu_{i}[k]$ have size $M=10$ and are zero-mean
white Gaussian distributed with covariance matrices $R_{u,i}=\sigma_{u,i}^2I_M$, with $\sigma_{u,i}^2$ shown on the bottom side of Fig.
\ref{net}. The noise variables are assumed to be distributed according to a GMRF, whose statistical dependency graph is depicted through
the thick links in Fig. \ref{net}. Each thick link is also supported by a communication link so that the dependency graph can be seen as
a sub-graph of the communication graph. Since the dependency graph in Fig. \ref{net} is acyclic, we compute the precision matrix as in
(\ref{prec_cov_matrices}) with $c_{i,i}=\sigma^2=0.0157$ and $c_{i,j}=\sigma^2 \nu \exp(-\kappa \cdot d_{ij})$, where $d_{ij}$
is the Euclidean distance among nodes $i$ and $j$, $\nu<1$ is the nugget parameter, and $\kappa\geq0$ is a correlation coefficient.

\begin{figure}[t]
\centering
\includegraphics[width=8.5cm]{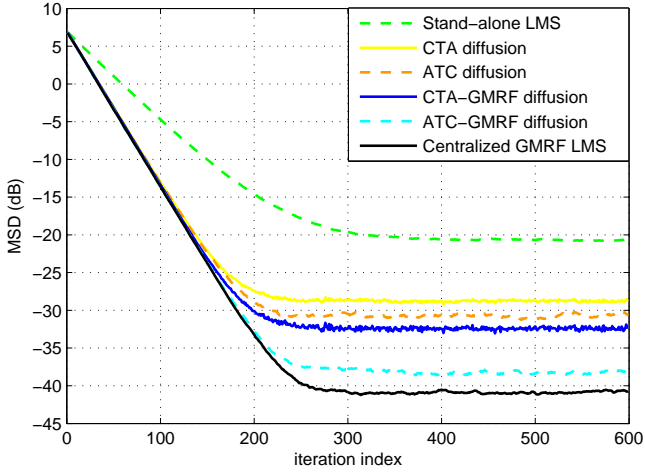}
  \caption{Network MSD versus iteration index, considering different algorithms.}\label{Comparison}
\end{figure}

\begin{figure}[t]
\centering
\includegraphics[width=8.5cm]{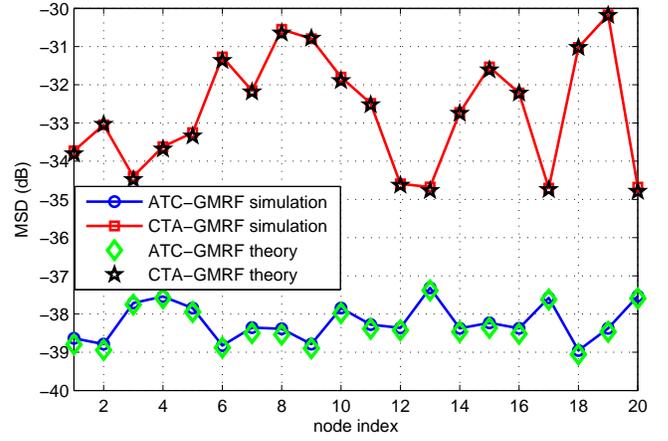}
  \caption{MSD versus node index, comparing theoretical results with numerical simulations.}\label{Theory}
\end{figure}

In this example, we aim to illustrate the potential gain offered by the proposed strategies in estimating a vector parameter embedded in a GMRF. To this goal, in Fig. \ref{Comparison} we show the learning behavior of 6 different strategies for adaptive filtering: stand alone LMS, CTA and ATC diffusion strategies from \cite{Cattivelli_Sayed}, the proposed CTA and ATC GMRF diffusion strategies in Tables 1 and 2, and the centralized LMS solution in (\ref{cent_LMS}). The parameters of the GMRF are $\nu=0.9$ and $\kappa=0.1$. The step-size of the GMRF diffusion strategies is equal to $3\times 10^{-4}$, whereas the step-sizes of the other algorithms are chosen in order to have the same convergence rate of the proposed strategies. We consider diffusion algorithms without measurement exchange, i.e. $\bQ=\bI$. Instead, the combination matrix $\bW$ in (\ref{combination_coefficients}) for the diffusion strategies is chosen such that each node simply averages the estimates from the neighborhood, i.e., $w_{ij}=1/|{\cal N}_i|$ for all $i$. As we can notice from Fig. \ref{Comparison}, thanks to the prior knowledge of the structure of the underlying dependency graph among the observations, the proposed ATC and CTA GMRF diffusion strategies lead to a gain with respect to their agnostic counterparts. The ATC strategy outperforms the CTA strategy, as in the case of standard diffusion LMS \cite{Cattivelli_Sayed}. From Fig. \ref{Comparison}, we also notice the large gain obtained by the diffusion strategies with respect to stand-alone LMS adaptation. Furthermore, we can see how the performance of the ATC-GMRF diffusion strategy is very close to the LMS centralized solution in (\ref{cent_LMS}), which has full knowledge of all the network parameters and observations. To check the validity of the theoretical derivations in (\ref{MSD_k}), in Fig. \ref{Theory} we illustrate the behavior of the steady-state MSD of the ATC and CTA GMRF diffusion strategies, at each node in the network, comparing the theoretical values with simulation results. The MSD values are obtained by averaging over 100 independent simulations and over 200 samples after convergence. From Fig. \ref{Theory}, we can notice the good matching between theory and numerical results.

\begin{figure}[t]
\centering
\includegraphics[width=8.5cm]{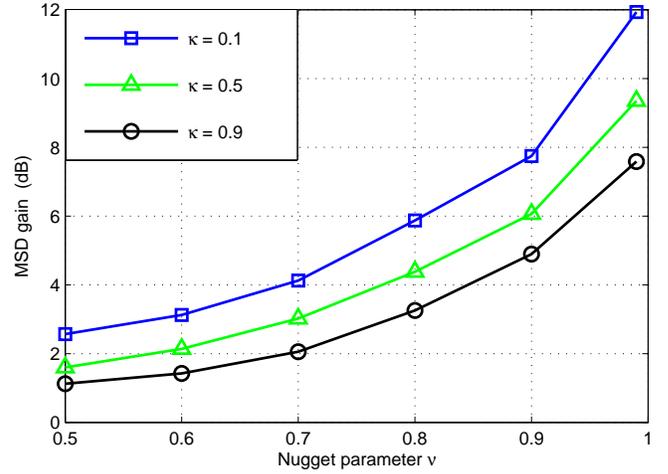}
  \caption{MSD gain versus $\nu$, for different values of $\kappa$.}\label{MSD_nugget}
\end{figure}

To assess the sensitivity of the proposed strategies to variations in the parameters describing the GMRF, in Fig. \ref{MSD_nugget},
we report the difference in dB between the steady-state network MSD of the ATC (from \cite{Cattivelli_Sayed}) and ATC-GMRF (table 1)
diffusion algorithms (i.e., the gain in terms of MSD), versus the nugget parameter $\nu$, considering different values of the coefficient
$\kappa$. The results are averaged over 100 independent realizations and over 200 samples after convergence. The parameters are the same of
the previous simulation and, for any pair $(\nu,\kappa)$, the step-sizes of the two algorithms are chosen in order to match their convergence
rate. As we can see from Fig. \ref{MSD_nugget}, as expected, the MSD gain improves by increasing the correlation among the observations, i.e.
by increasing the nugget parameter $\nu$ and reducing the coefficient $\kappa$.

\begin{figure}[t]
\centering
\includegraphics[width=8.5cm]{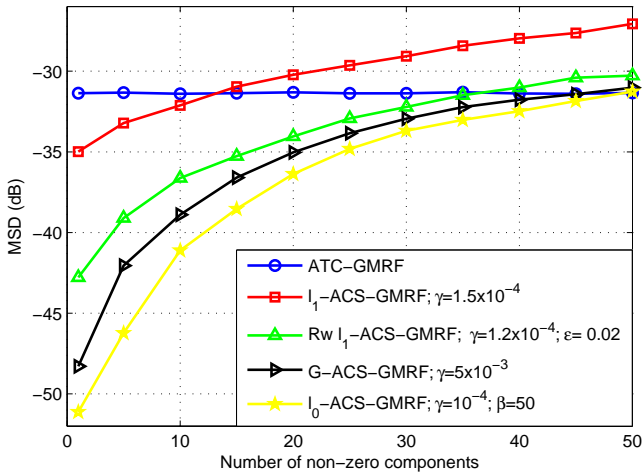}
  \caption{MSD versus number of non zero components of $\boldsymbol{\theta}_0$, considering different algorithms.}\label{MSD_deg}
\end{figure}

\vspace{.2cm}
{\it Numerical Example - Sparsity Recovery :} This example aims to show the steady-state performance for the sparse GMRF diffusion algorithms,
considering the different thresholding functions illustrated in Section V. The regressors $\bu_{i}[k]$ have size $M=50$ and are zero-mean
white Gaussian distributed with covariance matrices $R_{u,i}=\sigma_{u,i}^2I_M$, with $\sigma_{u,i}^2$ shown on the bottom side of Fig. \ref{net}.
In Fig. \ref{MSD_deg}, we report the steady-state network Mean Square Deviation (MSD), versus the number of non-zero components of the true
vector parameter (which are set to 1), for 5 different adaptive filters: the ATC-GMRF diffusion described in Table 1 (i.e., the sparsity agnostic GMRF diffusion
algorithm), the $\ell_1$-ACS GMRF diffusion, the Rw-$\ell_1$-ACS GMRF diffusion, the G-ACS GMRF diffusion, and the $\ell_0$-ACS GMRF diffusion,
which are described in Table 3 and by (\ref{Lasso}), (\ref{Rw_l1}), (\ref{Garotte}), and (\ref{l0}), respectively. The results are averaged over
100 independent experiments and over 200 samples after convergence. The step-sizes are chosen as $\mu_i=2.8\times10^{-4}$ for all $i$,
and the parameters of the GMRF are $\nu=0.9$ and $\kappa=0.1$. The combination matrix $\bW$  is chosen such that $w_{ij}=1/|{\cal N}_i|$ for
all $i$. The threshold parameters of the various strategies are available in Fig. \ref{MSD_deg}.
As we can see from Fig. \ref{MSD_deg}, when the vector is very sparse all the sparsity-aware strategies yield better steady-state performance
than the sparsity agnostic algorithm. The Rw-$\ell_1$, the garotte, and the $\ell_0$ estimators greatly outperform the lasso thanks to
the modified thresholding operations in (\ref{Rw_l1}), (\ref{Garotte}), and (\ref{l0}). When the vector is less sparse, the $\ell_1$-ACS GMRF
strategy performs worse than the sparsity agnostic algorithm due to the dominant effect of the bias introduced by the function in (\ref{Lasso}),
whereas the other strategies still lead to a positive gain. In particular, while in this example the Rw-$\ell_1$-ACS GMRF and the G-ACS GMRF
diffusion strategies perform worse than the sparsity agnostic ATC-GMRF diffusion algorithm if the number of non-zero components is larger than 37 and 45,
respectively, the $\ell_0$-ACS GMRF strategy leads always to a positive gain, thus matching the performance of the sparsity agnostic
strategy only when the vector $\boldsymbol{\theta}_0$ is completely non-sparse.

\begin{figure}[t]
\centering
\includegraphics[width=8.5cm]{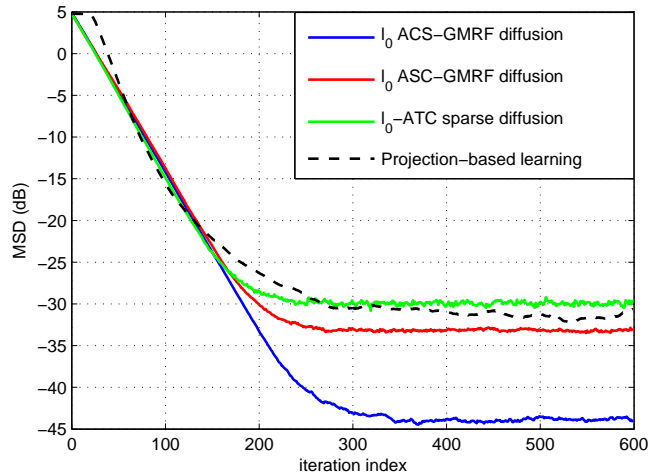}
  \caption{Network MSD versus iteration index, considering different algorithms.}\label{Comparison2}
\end{figure}

To compare the performance of the proposed strategies with other distributed, sparsity-aware, adaptive techniques available in the
literature, we illustrate the temporal behavior of the network MSD of four adaptive filters: the $\ell_0$-ACS GMRF described in Table 3
and by (\ref{l0}), the $\ell_0$-ASC GMRF described in Table 4 and by (\ref{l0}), the $\ell_0$-ATC sparse diffusion LMS from \cite{Liu-Li-Zhang},
 \cite{DiLorenzo-Sayed}, \cite{Dilorenzo-Barbarossa-Sayed2}, and the projection based sparse learning from \cite{Chouvardas-Slavakis-Kopsinis-Theodoridis}. The results are averaged
 over 100 independent experiments. We consider a vector parameter $\boldsymbol{\theta}_0$ with only 6 elements set equal to one,
which have been randomly chosen. The threshold parameters of the $\ell_0$-ACS GMRF (and $\ell_0$-ASC GMRF) are chosen such that $\gamma=10^{-4}$,
and $\beta=50$. The step-sizes, the combination matrix $\bW$, and the GMRF parameters are chosen as before. Using the same notation adopted in
\cite{Liu-Li-Zhang}, the parameters of the $\ell_0$ Sparse diffusion are $\rho=2\times10^{-3}$ and $\alpha=5$. Using the same notation adopted
in \cite{Chouvardas-Slavakis-Kopsinis-Theodoridis}, the parameters of the projection based filter are:
$\varepsilon=1.3\times\max_k(\sigma_{v,i})$; $\mu_n=0.06\times\mathcal{M}_n$; the radius of
the weighted $\ell_1$ ball is equal to $\|w^o\|_0=6$ (i.e., the correct sparsity level); $\tilde{\varepsilon}_n=0.02$; $\alpha=0.85$
for $k<160$ and $\alpha=0.65$ for $k>160$; the number of hyperslabs used per time update is equal to $q=20$.  From Fig. \ref{Comparison2},
we notice how the $\ell_0$-ACS GMRF algorithm outperforms all the other strategies. This is due to the
exploitation of the prior knowledge regarding the underlying GMRF and the adoption of the thresholding function in (\ref{l0}),
which gives powerful capabilities of sparsity recovery to the algorithm. As previously intuited in section V, ACS strategies outperform
ASC strategies thanks to the exploitation of the cooperation among nodes for noise reduction before the sparsification step.
At the same time, since in the ASC implementation each node transmits to its neighbors only the entries belonging to the estimated vector
support, the information exchange in the network is greatly reduced. Thus, ACS and ASC strategies constitute an interesting tradeoff
between performance and communication burden. These two algorithms have both linear complexity, i.e., $O(4M)$. At the same time,
the $\ell_0$ ATC sparse diffusion LMS from \cite{Liu-Li-Zhang}, \cite{DiLorenzo-Sayed}, \cite{Dilorenzo-Barbarossa-Sayed2}, has a linear
complexity too, i.e. $O(3M)$, whereas the projection-based method is more complex, i.e., $O(M (3+q+\log M))$, due to the presence of $q$
projections onto the hyperslabs and 1 projection on the weighted $\ell_1$ ball per iteration. This discussion further enlighten the good
features of the proposed strategies for distributed, adaptive and sparsity-aware estimation.

\begin{figure}[t]
\centering
\includegraphics[width=8.5cm]{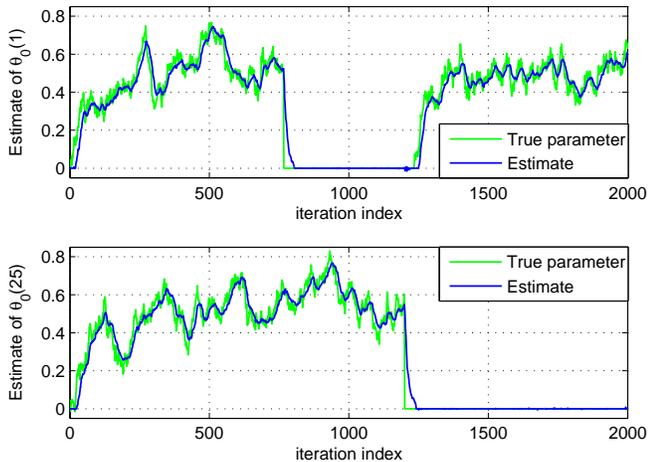}
  \caption{Example of tracking capability: Temporal behavior of the estimate of the first (top)
and twentyfifth (bottom) components of the time-varying vector $\boldsymbol{\theta}_0[k]$.}\label{tracking}
\end{figure}

\vspace{.2cm}
{\it Numerical Example - Tracking capability :} The aim of this example is to illustrate the tracking capability of the proposed strategies.
We consider the $\ell_0$-ACS GMRF described in Table 3 and by (\ref{l0}). In this example, the algorithm is employed to track a time-varying
parameter that evolves with time as $\boldsymbol{\theta}_0[k]=0.98\times\boldsymbol{\theta}_0[k-1]+\bs[k]$, where $\bs[k]$ is a Gaussian random
variable with mean $0.01\times\mathbf{1}_M$ and covariance matrix $4\times10^{-2}\bI$. In finite time intervals chosen at random,
the components of the vector parameter are set to zero. In Fig. \ref{tracking} we illustrate the behavior of the estimate of the first
and twentyfifth components of the time-varying vector $\boldsymbol{\theta}_0[k]$, superimposing also the true behavior of the parameters
for comparison purposes. The other parameters are the same of the previous simulation, except for the step-size that is set equal to $10^{-3}$.
As we can notice from Fig. \ref{tracking}, the algorithm tracks quite well the fluctuations of the parameter. Furthermore, thanks to the use of
the thresholding function in (\ref{l0}), the algorithm is also able to track sparsity in a very efficient manner, thus setting exactly to zero
the vector components that are found smaller than a specific threshold.

\section{Conclusions}

In this paper we have proposed distributed strategies for the online estimation of vectors over adaptive networks,
assuming the presence of spatially correlated measurements distributed according to a GMRF model. The proposed strategies
are able to exploit the underlying structure of the statistical dependency graph among the observations collected by the network
nodes at different spatial locations. A detailed mean square analysis has been carried out and confirmed by numerical simulations.
We have also illustrated how the proposed strategies can be extended by incorporating thresholding functions, which improve the performance
of the algorithms under sparsity of the vector parameter to be estimated. Several simulation results illustrate the potential advantages
achieved by these strategies for online, distributed, sparse vector recovery. 

The proposed methods require the apriori knowledge of the structure of the dependency graph existing among the observations collected
at different nodes. In practical applications, this means that the precision matrix must be previously estimated by the sensor network,
using a sparse covariance selection method, see, e.g. \cite{Meinhausen-Buhlmann} and references therein. Then, once each node is informed about the local structure of the dependency graph defined by the precision matrix,
the network can run the proposed strategies in a fully distributed fashion. An interesting future extension of this  work might be to couple
the proposed algorithms with (possibly distributed) online methods for covariance selection. In this way a further layer of adaptation
would be added into the system, thus enabling the network to track also temporal variations in the spatial correlation among data.
This problem will be the tackled in a future publication.

\appendices

\section{Proof of Theorem 1}

Letting $\bH=\hat{\bP}_2^T\left(\bI-\bM\bD\right)\hat{\bP}_1^T$,  recursion (\ref{compact_Diffusion_Expected}) gives
\begin{eqnarray}\label{recursion_expected}
\mathbb{E}\tilde{\boldsymbol{\theta}}[k]=\bH^k\mathbb{E}\tilde{\boldsymbol{\theta}}[0]
\end{eqnarray}
where $\tilde{\boldsymbol{\theta}}[0]$ is the initial condition. As long as we can show that $\bH^k$ converge to zero as $k$ goes to infinity, then we would be able to conclude the convergence of $\mathbb{E}\tilde{\boldsymbol{\theta}}[k]$. To proceed, we call upon results from \cite{Takahashi,Chen-Sayed,SayedBC}.
Let $\bz=\mbox{\rm col}\{\bz_1,\bz_2,\ldots,\bz_N\}$ denote a vector that is obtained by stacking $N$ subvectors of size $M\times 1$ each (as is the case with $\tilde{\boldsymbol{\theta}}[k]$).  The block maximum norm of $\bz$ is defined as
\begin{equation}
\|\bz\|_{b,\infty}\;=\; \displaystyle \max_{1\leq i\leq N}\;\|\bz_i\|,
\end{equation}
where $\|\cdot\|$ denotes the Euclidean norm of its vector argument. Likewise, the induced block maximum norm of a block matrix $\bX$ with $M\times M$ block entries is defined as:
\begin{equation}
\displaystyle \|\bX\|_{b,\infty}\;=\ \max_{\bz\neq \mathbf{0}}\;\frac{\|\bX\bz\|_{b,\infty}}{\|\bz\|_{b,\infty}}.
\end{equation}
Now, since
\begin{equation}\label{conv_cond}
\|\bH^k\mathbb{E}\tilde{\boldsymbol{\theta}}_0\|_{\infty}\leq\|\bH\|_{b,\infty}^k\cdot\|\mathbb{E}\tilde{\boldsymbol{\theta}}[0]\|_{b,\infty},
\end{equation}
recursion (\ref{recursion_expected}) converges to zero as $i\rightarrow \infty$ if we can ensure that $\|\bH\|_{b,\infty}<1$. This condition is actually satisfied by (\ref{step_sizes}). To see this, we note that
\begin{align}\label{conv_cond}
\|\bH\|_{b,\infty}&\leq \|\bI-\bM\bD\|_{b,\infty}
\end{align}
since $\left\| \hat{\bP}_1^T\right\|_{b,\infty}=\left\| \hat{\bP}_2^T\right\|_{b,\infty}=1$ in view of the fact that $\hat{\bP}_1$ and $\hat{\bP}_2$ are left-stochastic matrices \cite{Takahashi}. Therefore, to satisfy $\|\bH\|_{b,\infty}<1$, it suffices to require
\begin{equation}\label{conv_cond2}
\|\bI-\bM\bD\|_{b,\infty}<1.
\end{equation}
Now, we recall a result from \cite{Chen-Sayed} on the block maximum norm of a block diagonal and Hermitian matrix $\bX$ with $M\times M$ blocks $\{\bX_i\}$, which states that
\begin{equation}\label{conv_cond3}
\|\bX\|_{b,\infty}=\max_{i=1,\ldots,N}\rho(\bX_i)
\end{equation}
with $\rho(\bU)$ denoting the spectral radius of the Hermitian matrix $\bU$. Thus, since $\bM$ is diagonal, condition (\ref{conv_cond2}) will hold if the matrix $\bI-\bM\bD$ is stable. Using (\ref{perf_matrices3}), we can easily verify that this condition is satisfied for any step-sizes satisfying (\ref{step_sizes}), as claimed before. This concludes the proof of the theorem.

\section{Proof of Theorem 2}

Letting $\br={\rm vec}(\hat{\bP}_2^T\bM\bG^T\bM\hat{\bP}_2)$, recursion (\ref{weighted_norm3}) leads to:
\begin{eqnarray}\label{MSDi4}
\mathbb{E}\|\tilde{\boldsymbol{\theta}}[k]\|^2_{\textrm{vec}^{-1}(\boldsymbol{\sigma})}= \mathbb{E}\|\tilde{\boldsymbol{\theta}}[0]\|^2_{\textrm{vec}^{-1}(\boldsymbol{ F}^k\boldsymbol{\sigma})}+\br^T\sum_{l=0}^{k-1}\bF^l\boldsymbol{\sigma}
\end{eqnarray}
where $\mathbb{E}\|\tilde{\boldsymbol{\theta}}[0]\|^2$ is the initial condition. We first note that if $\bF$ is stable, $\bF^k \rightarrow \mathbf{0}$ as $k\rightarrow\infty$. In this way, the first term on the RHS of (\ref{MSDi4}) vanishes asymptotically. At the same time, the convergence of the second term on the RHS of (\ref{MSDi4}) depends only on the geometric series of matrices $\sum_{l=0}^{\infty}\bF^l$, which is known to be convergent to a finite value if the matrix $\bF$ is a stable matrix $\cite{Horn-Johnson}$. In summary, since both the first and second terms on the RHS of (\ref{MSDi4}) asymptotically converge to finite values, we conclude that $\mathbb{E}\|\tilde{\boldsymbol{\theta}}[k]\|^2_{\boldsymbol{\sigma}}$ will converge to a steady-state value, thus completing our proof.

\section{Proof of Theorem 3}

We will carry out the proof for the ACS strategy in (\ref{ACS diffusion}). The proof for the ASC strategy follows from straightforward modifications. Following the arguments in Section IV, we define the vectors $\tilde{\boldsymbol{\theta}}_{i}[k]=\boldsymbol{\theta}_0-\boldsymbol{\theta}_{i}[k]$, $\tilde{\boldsymbol{\zeta}}_{i}[k]=\boldsymbol{\theta}_0-\boldsymbol{\zeta}_{i}[k]$, and the network vectors:
\begin{eqnarray}\label{err_vectors}
\tilde{\boldsymbol{\theta}}[k]=\begin{bmatrix} \tilde{\boldsymbol{\theta}}_{1}[k] \\ \vdots \\ \tilde{\boldsymbol{\theta}}_{N}[k]  \end{bmatrix},\hspace{.1cm}
\tilde{\boldsymbol{\zeta}}[k]=\begin{bmatrix} \tilde{\boldsymbol{\zeta}}_{1}[k] \\ \vdots \\ \tilde{\boldsymbol{\zeta}}_{N}[k]  \end{bmatrix},\hspace{.1cm}
\boldsymbol{\zeta}[k]=\begin{bmatrix}\boldsymbol{\zeta}_{1}[k] \\ \vdots \\ \boldsymbol{\zeta}_{N}[k]
\end{bmatrix}
\end{eqnarray}
Then, the evolution of the error vector $\tilde{\boldsymbol{\theta}}[k]$ can be written as
\begin{align}\label{err_recursion_ACS}
\tilde{\boldsymbol{\theta}}[k]=\bI_N\otimes\boldsymbol{\theta}_0-\bT_\gamma(\boldsymbol{\zeta}[k])
\end{align}
The thresholding functions in (\ref{Lasso})-(\ref{l0}) can be cast as
\begin{align}\label{fx}
\bT_\gamma(\bx)=\bx+\bff(\bx), \quad\quad \|\bff(\bx)\|\leq c_1,
\end{align}
with $c_1=\gamma \sqrt{M}$ for (\ref{Lasso})-(\ref{Garotte}) and $c_1=\gamma\beta \sqrt{M}$ for (\ref{l0}). Then, substituting (\ref{fx}) in (\ref{err_recursion_ACS}), we have
\begin{align}\label{err_recursion_ACS_2}
\tilde{\boldsymbol{\theta}}[k]\;=\;&\tilde{\boldsymbol{\zeta}}[k]-\bff(\boldsymbol{\zeta}[k]) \nonumber\\
\;=\; &\bH[k] \tilde{\boldsymbol{\theta}}[k-1]- \hat{\bW}^T\bM\bg[k]-\bff(\boldsymbol{\zeta}[k])
\end{align}
with $\bH[k]=\hat{\bW}^T(\bI-\bM\bD[k])$ because, for the ACS strategy in (\ref{ACS diffusion}), we have $\hat{\bP}_2=\hat{\bW}=\bW\otimes\bI_N$ and $\hat{\bP}_1=\bI$. Taking the expectation of both terms in (\ref{err_recursion_ACS_2})
and letting $\bH=\mathbb{E}\bH[k]=\hat{\bW}^T(\bI-\bM\bD)$, the recursion can be cast as
\begin{align}\label{err_recursion_ACS_3}
\mathbb{E}\tilde{\boldsymbol{\theta}}[k]=\bH^k \mathbb{E}\tilde{\boldsymbol{\theta}}[0]-\sum_{l=0}^{k-1}\bH^l \cdot \mathbb{E}\{\bff(\boldsymbol{\zeta}[k-l])\}
\end{align}
Taking the block maximum norm of $\mathbb{E}\tilde{\boldsymbol{\theta}}[k]$ in (\ref{err_recursion_ACS_3}) and exploiting the boundness of function $\bff(\cdot)$, we have
\begin{align}\label{err_recursion_ACS_4}
\|\mathbb{E}\tilde{\boldsymbol{\theta}}[k]\|_{b,\infty}\leq\|\bH\|_{b,\infty}^k \|\mathbb{E}\tilde{\boldsymbol{\theta}}[0]\|_{b,\infty}+c_2\sum_{l=0}^{k-1}\|\bH\|_{b,\infty}^l
\end{align}
where $0<c_2<\infty$.  The right-hand side of (\ref{err_recursion_ACS_4}) converges as $k\rightarrow\infty$ to a fixed value if $\|\bH\|_{b,\infty}<1$. As shown in Appendix A, this condition is verified by choosing the step-sizes in order to satisfy (\ref{step_sizes}). This proves the stability in the mean of the ACS strategy (\ref{ACS diffusion}).

To prove the stability of the ACS strategy (\ref{ACS diffusion}) in the mean-square sense, using the same notation of Section IV.B and letting $\br={\rm vec}(\hat{\bP}_2^T\bM\bG^T\bM\hat{\bP}_2)$, we have from (\ref{err_recursion_ACS_2}) that
\begin{eqnarray}\label{weighted_norm4}
\mathbb{E}\|\tilde{\boldsymbol{\theta}}[k]\|^2_{\boldsymbol{\Sigma}}=\mathbb{E}\|\tilde{\boldsymbol{\theta}}[k-1]\|^2_{\textrm{vec}^{-1}(\boldsymbol{ F}^k\boldsymbol{\sigma})}+\br^T\boldsymbol{\sigma}
+f_2(\tilde{\boldsymbol{\theta}}[k-1])
\end{eqnarray}
where
\begin{eqnarray}\label{f2}
f_2(\tilde{\boldsymbol{\theta}}[k-1])=\mathbb{E}\|\bff(\boldsymbol{\zeta}[k])\|^2_{\boldsymbol{\Sigma}}-\mathbb{E}\{2\bff(\boldsymbol{\zeta}[k])^T\boldsymbol{\Sigma}\bH\tilde{\boldsymbol{\theta}}[k-1]\}
\end{eqnarray}
Since $\bff(\cdot)$ and $\mathbb{E}\tilde{\boldsymbol{\theta}}[k]$ are bounded by positive constants for any $k$, we have
$|f_2(\tilde{\boldsymbol{\theta}}[k-1])|<c_3$, with $0<c_3<\infty$. The positive constant $c_3$ can be related to the quantity $\br^T\boldsymbol{\sigma}$ in (\ref{weighted_norm4}) through some constant $\upsilon\in\mathbb{R}^+$, say, $c_3=\upsilon \br^T\boldsymbol{\sigma}$. Thus, from (\ref{weighted_norm4}), we can derive the upper bound
\begin{eqnarray}\label{weighted_norm5}
\mathbb{E}\|\tilde{\boldsymbol{\theta}}[k]\|^2_{\boldsymbol{\Sigma}}\leq\mathbb{E}\|\tilde{\boldsymbol{\theta}}[k-1]\|^2_{\textrm{vec}^{-1}(\boldsymbol{ F}^k\boldsymbol{\sigma})}+(1+\upsilon)\cdot\br^T\boldsymbol{\sigma},
\end{eqnarray}
which leads to the recursion
\begin{eqnarray}\label{MSDi5}
\mathbb{E}\|\tilde{\boldsymbol{\theta}}[k]\|^2_{\boldsymbol{\Sigma}}\leq \mathbb{E}\|\tilde{\boldsymbol{\theta}}[0]\|^2_{\textrm{vec}^{-1}(\boldsymbol{ F}^k\boldsymbol{\sigma})}+(1+\upsilon)\cdot\br^T\sum_{l=0}^{k-1}\bF^l\boldsymbol{\sigma}
\end{eqnarray}
where $\mathbb{E}\|\tilde{\boldsymbol{\theta}}[0]\|^2$ is the initial condition. Using the same arguments as in Appendix B, the right
hand side of (\ref{MSDi5}) converges to a fixed value if $\bF$ is a stable matrix. This proves the boundness of the quantity
$\mathbb{E}\|\tilde{\boldsymbol{\theta}}[k]\|^2_{\boldsymbol{\Sigma}}$ for all $k$ and, ultimately, the mean-square stability of the
ACS strategy (\ref{ACS diffusion}). This concludes the proof of the theorem.


\begin{thebibliography}{1}

\bibitem{Barb-Sard-Dilo}
S. Barbarossa, S. Sardellitti, and P. Di Lorenzo, ``Distributed Detection and Estimation in Wireless Sensor Networks,'' E-Reference Signal Processing, R. Chellapa and S. Theodoridis, Eds., Elsevier, 2013.

\bibitem{Bertsekas2}
D. Bertsekas, ``A new class of incremental gradient methods for least squares problems,'' \emph{SIAM Journal on Optimization}, vol. 7, no. 4, pp. 913--926, 1997.

\bibitem{Nedic-Bertsekas}
A. Nedic and D. Bertsekas, ``Incremental subgradient methods for nondifferentiable optimization,'' \emph{SIAM Journal on Optimization}, vol. 12, no. 1, pp. 109--138, 2001.


\bibitem{Lopes_Sayed2}
C. Lopes and A. H. Sayed, ``Incremental adaptive strategies over distributed networks,'' {\it IEEE Transactions on Signal Processing}, vol. 55, no. 8, pp. 4064--4077, 2007.

\bibitem{Li-Chambers-Lopes-Sayed}
L. Li, J. Chambers, C. Lopes, and A. H. Sayed, ``Distributed estimation over an adaptive incremental network based on the
affine projection algorithm,'' {\it IEEE Transactions on Signal Processing}, vol. 58, no. 1, pp. 151--164, 2009.

\bibitem{Karp}
R. M. Karp, ``Reducibility among combinational problems,'' {\it Complexity of Computer Computations} {\rm (R. E. Miller and J. W. Thatcher, eds.)}, pp. 85--104, 1972.

\bibitem{Schizas-Mateos-Giannakis}
I. D. Schizas, G. Mateos, and G. B. Giannakis, ``Distributed LMS for consensus-based in-network adaptive processing,'' {\it IEEE Trans. Signal Process.}, vol. 57, no. 6, pp. 2365--2382, Jun. 2009.

\bibitem{Lopes_Sayed}
C. G. Lopes and A. H. Sayed, ``Diffusion least-mean squares over adaptive networks: Formulation and performance analysis,'' {\it IEEE Transactions on Signal Processing}, vol. 56, no. 7, pp. 3122--3136, July 2008.

\bibitem{Cattivelli_Sayed}
F. S. Cattivelli and A. H. Sayed, ``Diffusion LMS strategies for distributed estimation,'' {\it IEEE Trans. on Signal Proc.}, vol. 58, pp. 1035--1048, 2010.

\bibitem{Tu-Sayed2}
S.-Y. Tu and A. H. Sayed, ``Diffusion strategies outperform consensus strategies for distributed estimation over adaptive networks,'' {\it IEEE Trans. Signal Processing}, vol. 60, no. 12, pp. 6217-6234, Dec. 2012.

\bibitem{Takahashi}
N. Takahashi, I. Yamada, and A. H. Sayed, ``Diffusion least-mean squares with adaptive combiners: Formulation and performance analysis,'' {\it IEEE Transactions on Signal Processing}, vol. 58, no. 9, pp. 4795--4810, Sep. 2010.

\bibitem{Chen-Sayed}
J. Chen and A. H. Sayed, ``Diffusion adaptation strategies for distributed optimization and learning over networks,''{\em IEEE Transactions on Signal Processing}, vol. 60, no. 8, pp. 4289-4305, 2012.

\bibitem{SayedBC}
A. H. Sayed, ``Diffusion adaptation over networks,'' in {\it Academic Press Library in Signal Processing}, vol. 3, R. Chellapa and S. Theodoridis, editors, pp. 323-454, Academic Press, Elsevier, 2013.

\bibitem{Cattivelli-Sayed3}
F. Cattivelli and A. H. Sayed, ``Modeling bird flight formations using diffusion adaptation,'' {\it IEEE Transactions on Signal Processing}, vol. 59, no. 5, pp. 2038-2051, May 2011.

\bibitem{Tu-Sayed}
S-Y. Tu and A. H. Sayed, ``Mobile adaptive networks,'' {\it IEEE J. Sel.Topics on Signal Processing}, vol. 5, no. 4, pp. 649-664, August 2011.

\bibitem{Chen-Zhao-Sayed}
J. Chen, X. Zhao, and A. H. Sayed, ``Bacterial motility via diffusion adaptation,'' {\it Proc. 44th Asilomar Conference on Signals, Systems and Computers}, Pacific Grove, CA, Nov. 2010.

\bibitem{Dilo-Barb-Sayed}
P. Di Lorenzo, S. Barbarossa, and Ali H. Sayed, ``Bio-Inspired Decentralized Radio Access based on Swarming Mechanisms over Adaptive Networks,'' {\it IEEE Transactions on Signal Processing}, Vol. 61, no. 12, pp. 3183-3197, 15 June 2013.

\bibitem{Dilorenzo-Barbarossa-Sayed2}
P. Di Lorenzo, S. Barbarossa, and A. H. Sayed, ``Distributed Spectrum Estimation for Small Cell Networks based on Sparse Diffusion Adaptation,'' {\it  IEEE Signal Processing Letters}, Vol. 20, no. 12, pp. 1261-1265, December 2013.

\bibitem{Chouvardas-Slavakis-Theodoridis}
S. Chouvardas, K. Slavakis, and S. Theodoridis, ``Adaptive robust distributed learning in diffusion sensor networks,'' {\it IEEE Transactions on Signal Processing}, vol. 59, no. 10, pp. 4692--4707, 2011.

\bibitem{Towfic-Chen-Sayed}
Z. Towfic, J. Chen and A. H. Sayed, ``Collaborative learning of mixture models using diffusion adaptation,'' in {\it Proc. IEEE Workshop on Machine Learning for Signal Processing}, Beijing, China, Sept. 2011.




\bibitem{Donoho}
D. Donoho, ``Compressed sensing,'' {\it IEEE Transactions on Information Theory}, vol. 52, no. 4, pp. 1289--1306, 2006.

\bibitem{Baraniuk}
R. Baraniuk, ``Compressive sensing,'' {\it IEEE Signal Processing Magazine}, vol. 25, pp. 21--30, March 2007.

\bibitem{Tibshirani}
R. Tibshirani, ``Regression shrinkage and selection via the LASSO,'' {\it J. Royal Statistical Society: Series B}, vol. 58, pp. 267--288, 1996.

\bibitem{Mateos-Bazerque-Giannakis}
G. Mateos, J. A. Bazerque, and G. B. Giannakis, ``Distributed sparse linear regression,'' {\it IEEE Transactions on Signal Processing}, vol 58, No. 10, pp. 5262--5276, Oct.  2010.

\bibitem{Bazerque-Giannakis}
J. A. Bazerque, and G. B. Giannakis, ``Distributed spectrum sensing for cognitive radio networks by exploiting sparsity,'' {\it IEEE Transactions on Signal Processing}, vol. 58, No. 3, pp. 1847-1862, March 2010.

\bibitem{Chen-Gu-Hero}
Y. Chen, Y. Gu, and A.O. Hero, ``Sparse LMS for system identification,'' in {\it Proc. IEEE International Conference on Acoustics, Speech, and Signal Processing}, pp. 3125--3128, Taipei, May 2009.

\bibitem{Gu-Jin-Mei}
Y. Gu, J. Jin, and S. Mei, ``$\ell_0$ norm constraint lms algorithm for sparse system identification,'' {\em IEEE Signal Processing Letters}, vol. 16, no. 9, pp. 774--777, 2009.

\bibitem{Angelosante-Bazerque-Giannakis}
D. Angelosante, J.A. Bazerque, and G.B. Giannakis, ``Online adaptive estimation of sparse signals: where RLS meets the $\ell_{1}$-norm,'' {\it IEEE Trans. on Signal Processing}, vol. 58, no. 7, pp. 3436--3447, July, 2010.

\bibitem{Babadi-Kalouptisidis-Tarokh}
B. Babadi, N. Kalouptsidis, and V. Tarokh, ``SPARLS: The sparse RLS algorithm,'' {\it IEEE Transactions on Signal Processing}, vol. 58, no. 8, pp. 4013--4025, Aug., 2010.

\bibitem{Kopsinis-Slavakis-Theodoridis}
Y. Kopsinis, K. Slavakis, and S. Theodoridis, ``Online sparse system identification and signal reconstruction using projections onto weighted $\ell_{1}$ balls,'' {\it IEEE Transactions on Signal Processing}, vol. 59, no. 3, pp. 936--952, March, 2010.

\bibitem{Slavakis-Kopsinis-Theodoridis-McLaughlin}
K. Slavakis, Y. Kopsinis, S. Theodoridis, and S. McLaughlin, ``Generalized thresholding and online sparsity-aware learning in a union of subspaces,'' {\it IEEE Transactions on Signal Processing}, vol. 61, no. 15, pp. 3760-3773, 2013.


\bibitem{Mota-Xavier-Aguiar-Puschel}
J. Mota, J. Xavier, P. Aguiar, and M. Puschel, ``Distributed basis pursuit,''  {\it IEEE Transactions on Signal Processing}, vol. 60, no. 4, April, 2012.

\bibitem{Chouvardas-Slavakis-Kopsinis-Theodoridis}
S. Chouvardas, K. Slavakis, Y. Kopsinis, S. Theodoridis, ``A sparsity-promoting adaptive algorithm for distributed learning,'' {\em IEEE Transactions on Signal Processing}, vol. 60, no. 10, pp. 5412--5425, Oct. 2012.

\bibitem{Liu-Li-Zhang}
Y. Liu, C. Li and Z. Zhang, ``Diffusion sparse least-mean squares over networks,'' {\em IEEE Transactions on Signal Processing}, vol. 60, no. 8, pp. 4480--4485, Aug. 2012.

\bibitem{Dilorenzo-Barbarossa-Sayed}
P. Di Lorenzo, S. Barbarossa, and A. H. Sayed, ``Sparse diffusion LMS for distributed adaptive estimation,'' {\it Proc. IEEE Int. Conf. on Acoustics, Speech, and Signal Proc.}, pp. 3281-3284, Kyoto, Japan, March 2012.

\bibitem{DiLorenzo-Sayed}
P. Di Lorenzo and A. H. Sayed, ``Sparse distributed learning based on diffusion adaptation,'' {\it IEEE Transactions on Signal Processing}, vol. 61, no. 6, pp. 1419--1433, 15 March 2013.

\bibitem{Sardellitti-Barbarossa}
S. Sardellitti and S. Barbarossa, ``Distributed RLS estimation for cooperative sensing in small cell networks,'' {\it IEEE Inter. Conf. on Acoustics, Speech and Signal Process.}, Vancouver, Canada, 2013.

\bibitem{Liu-Liu-Li}
Z. Liu, Y. Liu and C. Li, ``Distributed sparse recursive least squares over networks,'' {\em IEEE Transactions on Signal Processing}, vol. 62, no. 6, pp. 1386--1395, March 2014.


\bibitem{Lauritzen-book} S. L. Lauritzen,\emph{ Graphical Models}, Oxford University Press, 1996.

\bibitem{Wiesel-Hero}
A. Wiesel and A. O. Hero III, ``Distributed Covariance Estimation in Gaussian Graphical Models,''IEEE Transactions On Signal Processing, Vol. 60, No. 1, January 2012

\bibitem{Meinhausen-Buhlmann}
N. Meinshausen and P. Buhlmann, ``High-dimensional graphs and variable selection with the Lasso,''
{\it Annals of Statistics}, vol. 34, no. 3, pp. 1436-1462, 2006.

\bibitem{Dog-Liu}
A. Dogandzic and K. Liu, ``Decentralized random-field estimation for sensor networks using quantized spatially correlated data and fusion-
center feedback,'' \emph{IEEE Trans. Signal Process.}, vol. 56, no. 12, pp. 6069--6085, Dec. 2008.

\bibitem{Sudd-Wain-Will}
E. B. Sudderth, M. J. Wainwright, and A. S. Willsky, ``Embedded trees: Estimation of Gaussian processes on graphs with cycles,'' IEEE Transactions
on Signal Processing, vol. 52, no. 11, pp. 3136--3150, Nov. 2004.

\bibitem{Delouille-Neelami-Baraniuk}
V. Delouille, R. N. Neelamani, and R. G. Baraniuk, ``Robust distributed estimation using the embedded subgraphs algorithm,''
\emph{IEEE Trans. Signal Process.}, vol. 54, no. 8, pp. 2998--3010, Aug. 2006.

\bibitem{Chandr-Johnson-Willsky}
V. Chandrasekaran, J. K. Johnson, and A. S. Willsky, ``Estimation in Gaussian graphical models using tractable subgraphs: A walk-sum
analysis,'' IEEE Trans. Signal Process., vol. 56, no. 5, pp. 1916--1930, May 2008.

\bibitem{Fang-Li}
J. Fang and H. Li, `` Distributed estimation of Gauss-Markov random fields with one-bit quantized data,'' {\em IEEE Signal Processing Letters}, Vol. 17, no. 5, pp. 449--452, May 2010.

\bibitem{Anandkumar-Tong-Swami}
A. Anandkumar, L. Tong, A. Swami, ``Detection of Gauss-Markov Random Fields With Nearest-Neighbor Dependency,'' {\em  IEEE Trans. on Information Theory}, Vol. 55, pp. 816--827, Feb. 2009.

\bibitem{Kay}
S. M. Kay, {\it Fundamentals of Statistical Signal Processing: Estimation Theory}, Prentice Hall, 1993.

\bibitem{Bertsekas}
D. P. Bertsekas and J. N. Tsitsiklis, {\it Parallel and distributed computation: numerical methods}, Belmont, MA, Athena Scientific, 1997.

\bibitem{Nevel}
M. Nevelson and R. Hasminskii, \textit{Stochastic approximation and recursive estimation}, Providence, Rhode Island: American Math. Soc., 1973.

\bibitem{Candes-Wakin-Boyd}
E.J. Candes, M. Wakin, and S. Boyd, ``Enhancing sparsity by reweighted $\ell_{1}$ minimization,'' {\it Journal of Fourier Analysis and Applications}, vol. 14, pp.877--905, 2007.

\bibitem{Yuan-Lin}
M. Yuan, Y. Lin, ``On the non-negative garrotte estimator,'' {\it J. Royal Statist. Soc.}, vol. 69, pp. 143–161, 2007.

\bibitem{DiLo_Barb}
P. Di Lorenzo and S. Barbarossa, ``A bio-inspired swarming algorithm for decentralized access in cognitive radio,'' {\it IEEE Trans. on Signal Processing}, vol. 59, no. 12, pp. 6160--6174, December 2011.

\bibitem{Kushner-Yin}
H. J. Kushner and G. G. Yin, {\it Stochastic Approximation Algorithms and Applications}, New York: Springer-Verlag, 1997.

\bibitem{Sayed}
A. H. Sayed, {\it Adaptive Filters}, Wiley, NJ, 2008.

\bibitem{Graham}
R. L. Graham, D. E. Knuth and O. Patashnik, {\it Concrete Mathematics: A Foundation for Computer Science}, 2nd ed., Addison-Wesley, 1994.

\bibitem{Kailath-Sayed}
T. Kailath, A. H. Sayed, and B. Hassibi, {\it Linear Estimation}, Englewood
Cliffs, NJ: Prentice-Hall, 2000.

\bibitem{Horn-Johnson}
R. Horn and C. Johnson, {\it Matrix Analysis}, Cambridge university press, 2005.



\end{thebibliography}
\end{document}